\documentclass[10pt,final,journal,twocolumn]{IEEEtran}
\ifCLASSINFOpdf

\else
\usepackage{algorithm}
\usepackage{booktabs}

\fi
\usepackage{cite}
\usepackage[cmex10]{amsmath}
\usepackage{amsthm}
\usepackage{stfloats}
\usepackage{multicol,multienum}
\usepackage{multirow}
\usepackage{mdframed}
\usepackage{amssymb}
\usepackage{url}
\usepackage{amsmath}
\usepackage{xcolor}
\usepackage{graphicx}
\usepackage{subfigure}
\usepackage{caption}
\usepackage{amsmath}
\usepackage{amsfonts,amssymb}
\usepackage{bbm}
\usepackage[T1]{fontenc}
\usepackage{algorithm}
\usepackage{algpseudocode}
\usepackage{ulem}
\usepackage{balance}
\usepackage[linesnumbered, ruled]{}
\makeatletter
\renewcommand{\maketag@@@}[1]{\hbox{\m@th\normalsize\normalfont#1}}%
\makeatother

\usepackage{caption}
\begin{document}
\hyphenation{op-tical net-works semi-conduc-tor}
\title{Trajectory Design for Fairness Enhancement in Movable Antennas-Aided Communications}
\author{Guojie Hu, Qingqing Wu, Lipeng Zhu, Kui Xu, Guoxin Li and Tong-Xing Zheng\vspace{-1.2em}
\thanks{
%
Guojie Hu is with the College of Communication Engineering, Rocket Force University of Engineering, Xi'an 710025, China (lgdxhgj@sina.com). Qingqing Wu is with the Department of Electronic Engineering, Shanghai Jiao Tong University, Shanghai 200240, China. Lipeng Zhu is with the State Key Laboratory of CNS/ATM and the School of Interdisciplinary Science, Beijing Institute of Technology, Beijing 100081, China. Kui Xu and Guoxin Li are with the College of Communications Engineering, Army Engineering University of PLA, Nanjing 210007, China. Tong-Xing Zheng is with the School of Information and Communications Engineering, Xi'an Jiaotong University, Xi'an 710049.
}
}
\IEEEpeerreviewmaketitle
\maketitle
\begin{abstract}
Through adaptive antenna repositioning, the movable antenna (MA) technology enables on-demand reconfiguration of wireless channels, thereby creating an additional spatial degree of freedom in improving communication performance. This paper investigates a multiuser uplink communication system aided by MAs, where a base station (BS) equipped with multiple MAs serves multiple single-antenna users. Specifically, given that an optimized array geometry cannot guarantee rate fairness, we focus on designing antenna trajectory at the BS to maximize the minimum achievable rate among all users over a finite time period. The resulting optimization problem is fundamentally challenging to solve due to the continuous-time nature. To address it, we first examine an ideal case with infinitely fast MA movement and demonstrate that the relaxed problem can be optimally solved via the Lagrangian dual method. The obtained trajectory solution reveals that the BS should employ a finite set of MA deployment patterns, each allocated an optimal time duration. Building on this, we then study the general case with limited MA movement speed and propose a heuristic trajectory design inspired by the optimal patterns identified in the ideal scenario. Several insights are also gained by examining the simplified special case. Finally, numerical results are provided to validate the effectiveness of the proposed designs compared to competitive benchmarks.
\end{abstract}
\begin{IEEEkeywords}
Movable antenna, multiuser uplink, antenna trajectory design, fairness.
\end{IEEEkeywords}

\IEEEpeerreviewmaketitle
\vspace{-10pt}
\section{Introduction}
Multiple-input multiple-output (MIMO) technology has been a pivotal enabler for modern wireless communication systems, delivering substantial improvements in capacity, reliability, and spectral efficiency compared to their single-antenna counterparts \cite{MIMO1}. However, conventional MIMO typically relies on fixed-position antenna (FPA) arrays \cite{MIMO2}, where antenna positions are static and unable to adapt, resulting in rigid wireless channel characteristics. Consequently, in unfavorable propagation conditions, FPA-based MIMO systems cannot unleash their full potential of spatial multiplexing and beamforming, leading to suboptimal system performance.

A natural idea is to further increase the number of antennas and radio-frequency (RF) units at transceivers to enhance channel conditions \cite{Massive_MIMO2}. However, this approach substantially raises hardware costs and energy consumptions. This prompts a critical question: Is there a way to improve system performance without increasing hardware complexities? In this context, the movable antenna (MA) technology has emerged as a promising paradigm to overcome the limitations of traditional FPAs \cite{Lipeng1, Lipeng2, Lipeng3, Lipeng4,gao2026two}. Instead of simply deploying more antennas or RF units, MAs enhance communication performance by adaptively reconfiguring antenna positions within a specific spatial region spanning from several wavelengths to tens of meters, leveraging technologies such as stepper motors \cite{Lipeng5, E_Lipeng}. Predictably, via flexible antenna position deployment, wireless channel conditions can be deliberately shaped into favorable configurations, thus improving spatial degrees of freedom (DoFs) and establishing a solid foundation for efficient beamforming.

 \begin{figure} [!t]
\centering
\includegraphics[width=8cm]{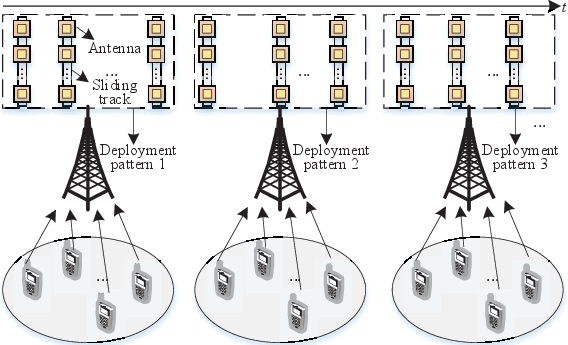}
\captionsetup{font=small}
\caption{Illustration of the considered system model.}
\label{fig:Fig1}
\end{figure}

Motivated by the potential advantages of the MA technology, numerous works have examined its performance across a variety of system scenarios. For instance, early studies established foundational channel models, such as the field-response model for narrowband single-input single-output (SISO) systems \cite{Lipeng1} and its extensions to wideband \cite{MA_wideband1, MA_wideband2}, MIMO \cite{MA_MIMO1, MA_MIMO2, MA_MIMO3, E_Lipeng2}, and near-field scenarios \cite{MA_near1}. The concept was later generalized to six-dimensional MA (6DMA), incorporating both position and orientation adjustments \cite{6DMA1, 6DMA2}. For point-to-point links, MA positioning has been shown to significantly improve the channel gain in SISO systems \cite{Lipeng1} and capacity in MIMO systems \cite{Lipeng2}. In multiuser networks, research has focused on uplink scenarios with single-MA users or MA-equipped base station (BS)\cite{MA_multiuser1, XZY_TWC, MA_multiuser3} and downlink beamforming with MA-equipped BS \cite{MA_multiuser4, MA_multiuser5, MA_MIMO2}, employing optimization methods ranging from gradient-based algorithms to two-timescale frameworks. The MA technology has also been applied to integrated sensing and communication (ISAC) \cite{ISAC1, ISAC2, MA-ISAC}, where antenna repositioning improves sensing accuracy while maintaining communication performance. Efficient channel acquisition strategies, including compressed sensing \cite{Compressed_Sensing, XZY_11} and tensor decomposition \cite{Tensor}, have been developed to estimate the continuous channel response. Furthermore, MA systems have been integrated with emerging technologies such as intelligent reflecting surfaces \cite{MA_IRS1, MA_IRS2,peng2025double,peng2025rotatable,peng2025single,MIS1,MIS2,MIS3}, physical layer security \cite{MA_PLS}, over-the-air computation \cite{MA_computing}, and non-orthogonal multiple access \cite{MA_NOMA}, demonstrating their versatility in next-generation wireless networks.

In this paper, similar to \cite{XZY_TWC}, we address the problem of minimum rate maximization in the MAs-aided multiuser uplink communication system, where the BS is equipped with multiple MAs and serves several single-antenna users over a given transmission period. To find an effective solution, a straightforward and conventional approach, such as the one employed in \cite{XZY_TWC}, was to adopt a single and static antenna deployment throughout the entire transmission period. \textbf{However, such scheme inevitably leads to significant unfairness in the achievable rate across users, since even a carefully optimized static antenna configuration cannot guarantee similar channel power and interference conditions of all users.} Consequently, certain users may experience persistently poor service quality, thereby undermining overall system fairness. Motivated by the deficiencies observed in \cite{XZY_TWC}, we in this paper propose a novel scheme aimed at enhancing fairness in MAs-aided multiuser uplink systems. Specifically, we focus on designing antenna trajectory at the BS to maximize the minimum achievable rate among all users over a finite transmission period, rather than relying on a single static antenna deployment. \textbf{This approach is grounded in the observation that dynamically adjusting antenna positions induces time-varying achievable rate across users, thereby creating an opportunity to balance their average rate via the time-sharing strategy.}

It should be emphasized that, even the trajectory design problem has already been studied in MAs-enabled systems \cite{MA_UAV1, MA_UAV2, MA_UAV3, MA_UAV4, MA_UAV5}, these works are fundamentally distinct from our focus. Specifically, i) \cite{MA_UAV1} focused on the macro-mobility of unmanned aerial vehicles (UAVs) coupled with the micro-positioning of MAs. In this study, the ``trajectory'' refers solely to the UAV's flight path, while the MA positions are still optimized only once and remain static throughout the transmission period; ii)  On the other hand, although \cite{MA_UAV2, MA_UAV3, MA_UAV4, MA_UAV5} investigated the MA trajectory design problem over the entire period under various scenarios, they primarily rely on general-purpose optimization techniques such as successive convex approximation (SCA) to obtain suboptimal solutions, offering limited structural insights to guide practical trajectory planning. Specifically, \cite{MA_UAV2, MA_UAV3, MA_UAV4, MA_UAV5} failed to address a critical question that must be carefully considered in antenna trajectory design: \textbf{Do antennas actually need to move frequently throughout the entire period? If so, frequent repositioning would incur substantial power consumption, which is a particularly undesirable feature for future low-power green communication systems. If not, then what is the optimal antenna movement strategy?}

This paper aims to fundamentally answer the above problem by focusing on the typical multiuser uplink communication system, with the main contributions summarized as follows.
\begin{itemize}
\item[$\bullet$] First, we adopt linear and optimal minimum mean-square error (MMSE) beamforming at the BS, which maximizes the received signal-to-interference-plus-noise ratio (SINR) of each user for any given antenna deployment. This allows the original minimum rate maximization problem to be transformed into an equivalent formulation that depends only on antenna positions at the BS, thereby establishing a tractable foundation for subsequent antenna trajectory optimization.

 \item[$\bullet$] Second, the resulting optimization problem is highly non-convex and involves continuous-time variables, making it challenging to solve directly. To gain essential insights, we first consider an ideal scenario with unlimited antenna movement speed, which relaxes the practical speed constraint. By solving this relaxed problem optimally via the Lagrangian dual method, we rigorously show that the BS only needs to adopt a finite set of antenna deployment patterns and allocate optimal time durations to each. This fundamental finding not only provides a performance upper bound but also reveals a low-movement and energy-efficient structure for practical MA trajectory design.

 \item[$\bullet$] Third, for the general case with limited antenna movement speed, we propose the Successive Stay-then-Move Trajectory (SSMT) scheme. This scheme sequentially visits the set of optimal deployment patterns identified from the ideal case. To enable continuous data reception throughout the entire transmission period, we rigorously prove that no signal coupling occurs during antenna repositioning. Leveraging this, the scheme jointly optimizes the switching trajectory, the time spent at each pattern, and the transition time between patterns, thereby maximizing the minimum average rate under a fixed total time period.

 \item[$\bullet$] Finally, through both analytical insights from a simplified case and comprehensive numerical simulations, we validate the efficiency and fairness enhancement of the proposed designs. The results demonstrate significant performance gains over the conventional static deployment and offer practical insights into the impacts of key system parameters such as the total antenna span, the antenna movement speed, the number of users and the channel estimation accuracy.
\end{itemize}

The rest of this paper is organized as follows. Section II introduces the system model and problem formulation. Section III investigates the ideal case with unlimited movement speed, derives the optimal set of deployment patterns via the Lagrangian dual method, and provides further insights through a simplified special case. Section IV proposes the practical SSMT scheme for the general case with limited speed. Its design is further illustrated and validated through a representative special case. Section V provides numerical results and performance analysis. Finally, Section VI concludes the paper.

\textit{Notations}: For a complex scalar $a$, ${\mathop{\rm Re}\nolimits} \left( a \right)$ denotes its real part. For a complex vector ${\bf{a}}$, ${{\bf{a}}^H}$, $\left\| {\bf{a}} \right\|$ and ${\left[ {\bf{a}} \right]_i}$ denote its conjugate transpose, Frobenius norm and the $i$-th element in ${\bf{a}}$, respectively. For a matrix ${\bf{A}}$, ${{\bf{A}}^H}$ and ${{\bf{A}}^{ - 1}}$ denote the conjugate transpose and the inverse, respectively. ${{\bf{I}}_N}$ is the identity matrix of size $N \times N$. ${\cal CN}(0,{\sigma ^2}{\bf{I}})$ denotes a circularly symmetric complex Gaussian distribution with zero mean and covariance matrix ${\sigma ^2}{\bf{I}}$. $ \odot $ denotes the Hadamard (element-wise) product, $ \otimes $ denotes the Kronecker product, and $\left\lfloor  \cdot  \right\rfloor $ denotes the floor function.

 \newcounter{mytempeqncnt}
\section{System Model and Problem Formulation}
\subsection{System Model}
As illustrated in Fig. 1, we consider the MAs-aided multiuser uplink communications, in which the BS equipped with $N \times M$ MAs serves $K$ single-antenna users during a finite time period, denoted by ${\cal T} \buildrel \Delta \over = [0,T]$. The BS employs a collective movement strategy for its antennas, in light of the fact that independently controlling each antenna unit to achieve flexible movement will result in the high hardware costs \cite{MA_Cross}. Specifically, the BS is configured with $M$ vertical sliding tracks on a 2D plane, each hosting an FPA-based uniform linear array of $N$ antennas spaced by half-wavelength.\footnote{Actually, similar to \cite{MA_Cross}, the BS can also employ the more advanced cross-link MA array consisting of vertical and horizontal sliding tracks to provide more flexible movement degree for each antenna. In this paper, we consider the simplified architecture in order to facilitate subsequent analysis.} These sliding tracks can independently move along the horizontal direction within a span of length $L$. Under this setup, we denote ${{x_m}}$ as the horizontal coordinate of the $m$-th sliding track relative to the reference point zero, with ${\cal M} \buildrel \Delta \over = \left\{ {1,...,M} \right\}$. Let ${\bf{x}} \buildrel \Delta \over = \left[ {{x_1},...,{x_M}} \right] \in {{\mathbb{R}}^{1 \times M}}$, where $0 \le {x_1} < ... < {x_M} \le L$ without loss of generality. In addition, by setting the vertical coordinate of the bottom antenna on each sliding track to zero, the vertical coordinates of $N$ antennas in each sliding track (arranging in ascending order from bottom to top) can be expressed as ${{\bf{y}}^{{\rm{FPA}}}} \buildrel \Delta \over = [{y_1},...,{y_N}] = \left[ {0:\lambda /2:(N - 1)\lambda /2} \right] \in {{\mathbb{R}}^{1 \times N}}$, where $\lambda $ is the signal wavelength.

Since the moving region of MAs (on the order of several wavelengths) is significantly smaller than the signal propagation distance, the far-field channel condition between the BS and users is assumed. Given the physical elevation angle of arrival (AoA) and azimuth AoA of user $k$ as ${\theta _k}$ and ${\phi _k}$, and considering the line-of-sight (LoS) propagation environment, the channel vector between the BS and user $k$ with the given antenna deployment pattern ${\bf{x}}$ can be expressed as
\begin{equation}
\begin{split}{}
{{\bf{h}}_k}({\bf{x}}) = \sqrt {{\beta _k}} {\left( {{\bf{f}}_k^{{\rm{hor}}}({\bf{x}})} \right)^H} \odot {\left( {{\bf{f}}_k^{{\rm{ver}}}({{\bf{y}}^{{\rm{FPA}}}})} \right)^H} \in {{\mathbb{C}}^{MN \times 1}},
\end{split}
\end{equation}
where ${{\beta _k}}$ is the large-scale fading coefficient and
\begin{equation} \nonumber
\begin{split}{}
{\bf{f}}_k^{{\rm{hor}}}({\bf{x}}) =&\left[ {f_k^{{\rm{hor}}}({{\left[ {\bf{x}} \right]}_1}),...,f_k^{{\rm{hor}}}({{\left[ {\bf{x}} \right]}_M})} \right] \in {{\mathbb{C}}^{1 \times M}},\\
{\bf{f}}_k^{{\rm{ver}}}({{\bf{y}}^{{\rm{FPA}}}}) =&\left[ {f_k^{{\rm{ver}}}({{\left[ {{{\bf{y}}^{{\rm{FPA}}}}} \right]}_1}),...,f_k^{{\rm{ver}}}({{\left[ {{{\bf{y}}^{{\rm{FPA}}}}} \right]}_N})} \right] \in {{\mathbb{C}}^{1 \times N}},
\end{split}
\end{equation}
with $f_k^{{\rm{hor}}}({\left[ {\bf{x}} \right]_m}) = {e^{j\frac{{2\pi }}{\lambda }{{\left[ {\bf{x}} \right]}_m}{\vartheta _k}}}$ and $f_k^{{\rm{ver}}}({\left[ {{{\bf{y}}^{{\rm{FPA}}}}} \right]_n}) = {e^{j\frac{{2\pi }}{\lambda }{{\left[ {{{\bf{y}}^{{\rm{FPA}}}}} \right]}_n}{\varphi _k}}}$, where ${\vartheta _k} \buildrel \Delta \over = \cos {\theta _k}\cos {\phi _k}$ and ${\varphi _k} \buildrel \Delta \over = \sin {\theta _k}$ denote the
virtual AoAs of user $k$.

Given ${\bf{x}}$ and the corresponding channel vectors $\left\{ {{{\bf{h}}_k}({\bf{x}})} \right\}_{k = 1}^K$, the received signal $\widehat {\bf{s}} \in {{\mathbb{C}}^{MN \times 1}}$ at the BS can be expressed as
\begin{equation}
\begin{split}{}
 \widehat {\bf{s}}= {{\bf{h}}_k}({\bf{x}})\sqrt {{P_k}} {s_k} + \sum\nolimits_{j = 1,j \ne k}^K {{{\bf{h}}_j}({\bf{x}})\sqrt {{P_j}} {s_j}}  + {\bf{n}},
\end{split}
\end{equation}
where $P_k$ and $s_k$ are the transmit power and information-bearing signal of user $k$, and ${\bf{n}} \sim {\cal CN}(0,{\sigma ^2}{{\bf{I}}_{MN}})$ is the additive white Gaussian noise at the BS, where ${\sigma ^2}$ is the average noise power. Then, by employing the receiving beamforming vector ${{\bf{w}}_k} \in {{\mathbb{C}}^{MN \times 1}}$ for detecting the signal transmitted by user $k$, the corresponding received SINR of user $k$ is
\begin{equation}
\begin{split}{}
{\gamma _k}({{\bf{w}}_k},{\bf{x}}) = \frac{{\overline {{P_k}} {{\left| {{\bf{w}}_k^H{{\bf{h}}_k}({\bf{x}})} \right|}^2}}}{{\sum\nolimits_{j = 1,j \ne k}^K {\overline {{P_j}} {{\left| {{\bf{w}}_k^H{{\bf{h}}_j}({\bf{x}})} \right|}^2} + 1} }},
\end{split}
\end{equation}
where $\overline {{P_k}}  = {P_k}/{\sigma ^2}$. Note that in (3), the SINR of user $k$ is only related to ${{\bf{w}}_k}$ and ${\bf{x}}$. Therefore, given ${\bf{x}}$, ${\gamma _k}({{\bf{w}}_k},{\bf{x}})$ can be maximized by the linear MMSE beamforming \cite{MA_multiuser1}, i.e.,
\begin{equation}
\begin{split}{}
{\bf{w}}_k^{{\rm{MMSE}}} = {\bf{A}}_k^{ - 1}({\bf{x}}){{\bf{h}}_k}({\bf{x}})/\left\| {{\bf{A}}_k^{ - 1}({\bf{x}}){{\bf{h}}_k}({\bf{x}})} \right\|,
\end{split}
\end{equation}
where ${{\bf{A}}_k}({\bf{x}}) = \sum\nolimits_{j = 1,j \ne k}^K {\overline {{P_j}} } {{\bf{h}}_j}({\bf{x}}){\bf{h}}_j^H({\bf{x}}) + {{\bf{I}}_{MN}} \in {{\mathbb{C}}^{MN \times MN}}$. Substituting ${\bf{w}}_k^{{\rm{MMSE}}}$ into (3), the SINR of user $k$ can be simplified into one function only related to ${\bf{x}}$ \cite{MA_multiuser1}, i.e.,
\begin{equation}
\begin{split}{}
{\gamma _k}({\bf{x}}) = \overline {{P_k}} {\widetilde \gamma _k}({\bf{x}}) = \overline {{P_k}} {\bf{h}}_k^H({\bf{x}}){\bf{A}}_k^{ - 1}({\bf{x}}){{\bf{h}}_k}({\bf{x}}).
\end{split}
\end{equation}

In the next, we will formulate the optimization problem based on (5), with the details provided below.
\subsection{Problem Formulation}
In this paper, the objective is to maximize the minimum rate among all users over the time period ${\cal T}$. To achieve this purpose, a fixed (static) antenna deployment pattern at the BS may be insufficient, as it tends to easily cause unfairness in terms of individual user rate. Therefore, the key idea of this paper is to design antenna trajectory of the BS over time $t \in {\cal T}$ (denoted by ${\left\{ {{{\bf{x}}^t}} \right\}_{t \in {\cal T}}}$), for adaptively adjusting the achievable rate of each user over time $t \in {\cal T}$. Through such dynamic adjustment, the average rate of different users can be predictably balanced over the entire period.

Given the antenna trajectory ${\left\{ {{{\bf{x}}^t}} \right\}_{t \in {\cal T}}}$, the average rate of user $k$ over the time period ${\cal T}$ (in bits/s/Hz) can be derived as
\begin{equation}
\begin{split}{}
\overline {{r_k}}  = \frac{1}{T}\int_0^T {{{\log }_2}\left( {1 + {\gamma _k}({{\bf{x}}^t})} \right)dt} .
\end{split}
\end{equation}

Aiming to maximize the minimum average rate among all users, the optimization problem can be formulated as
 \begin{align}
&({\rm{P1}}):{\rm{  }}\mathop {\max }\limits_{{{{\left\{ {{{\bf{x}}^t}} \right\}}_{t \in {\cal T}}}}} \ \mathop {\min }\limits_{k \in {\cal K}} \left\{ {\overline {{r_k}} } \right\}  \tag{${\rm{7a}}$}\\
{\rm{              }}&{\rm{s.t.}} \ \ \left| {d{{\left[ {{{\bf{x}}^t}} \right]}_m}/dt} \right| \le {V_{\max }},\forall t \in {\cal T},m \in {\cal M},\tag{${\rm{7b}}$}\\
 &\ \ \ \ \ {\rm{        }}{\left[ {{{\bf{x}}^t}} \right]_m} - {\left[ {{{\bf{x}}^t}} \right]_p} \ge {d_{\min }},\forall t \in {\cal T},m,p \in {\cal M},m > p,\tag{${\rm{7c}}$}\\
 &\ \ \ \ \ {\left[ {{{\bf{x}}^t}} \right]_m} \in [0,L],\forall t \in {\cal T},m \in {\cal M}, \tag{${\rm{7d}}$}
\end{align}
where ${\cal K} \buildrel \Delta \over = \left\{ {1,...,K} \right\}$, the constraint in (7b) indicates that the moving speed of each vertical sliding track does not exceed ${V_{\max }}$ at any moment, the constraint in (7c) implies that the horizontal distance between any two vertical sliding tracks at any moment should be greater than $d_{{\rm{min}}}$ to avoid coupling effects, and the constraint in (7d) shows that the horizontal coordinate of any vertical sliding track should lie in the interval $[0,L]$ in the entire period.

Note that the number of optimization variables in (P1) are infinite due to the continuous time $t$. In addition, the objective function of (P1) involves the complex structure with respect to (w.r.t.) ${{{\bf{x}}^t}}$, as reflected in (5). These two aspects cause huge obstacles for optimally solving (P1).

Next, we first consider an ideal scenario in which the movement speed of each vertical sliding track is unlimited, thereby solving a relaxed version of the problem to establish a performance upper bound. Building on this relaxation, we further propose one heuristic scheme to sub-optimally solve the original problem (P1) under the practical constraint of finite movement speed.

\section{Minimum Rate Maximization Without Antenna Movement Speed Constraint}
In this section, we first consider a relaxed problem by ignoring the movement speed constraint of each sliding track in (7b), leading to
 \begin{align}
&({\rm{P2}}):{\rm{  }}\mathop {\max }\limits_{{{{\left\{ {{{\bf{x}}^t}} \right\}}_{t \in {\cal T}}}}} \ \mathop {\min }\limits_{k \in {\cal K}} \left\{ {\overline {{r_k}} } \right\}  \tag{${\rm{8a}}$}\\
{\rm{              }}&{\rm{s.t.}} \ \ (7{\rm{c}}), (7{\rm{d}}).\tag{${\rm{8b}}$}
\end{align}

Note that the relaxed problem (P2) corresponds to the scenario where $V_{\rm{max}}$ is sufficiently large, or equivalently, the time period $T$ is sufficiently long (as will be explained in detail later). By introducing a slack variable $r$, problem (P2) can be equivalently formulated as
 \begin{align}
&({\rm{P2.1}}):{\rm{  }}\mathop {\max }\limits_{{{{\left\{ {{{\bf{x}}^t}} \right\}}_{t \in {\cal T}}}}, r } \ r  \tag{${\rm{9a}}$}\\
{\rm{              }}&{\rm{s.t.}} \ \ \ \ \overline {{r_k}}  \ge r,\forall k \in {\cal K},\tag{${\rm{9b}}$}\\
 &\ \ \ \ \ \ \ (7{\rm{c}}), (7{\rm{d}}).\tag{${\rm{9c}}$}
\end{align}

\subsection{Solving (P2.1) via Lagrange Dual Method}
Problem (P2.1) is still highly non-convex, however, it satisfies the so-called time-sharing condition \cite{Xujie_UAV}. This indicates that the strong duality holds between (P2.1) and its Lagrange dual problem. Therefore, problem (P2.1) can be solved with the Lagrange dual method \cite{Xujie_UAV}.

Specifically, let ${\mu _k} > 0$, $k = 1,...,K$, denote the dual variable associated with the constraint in (9b) for user $k$. The Lagrangian associated with (P2.1) can be expressed as
\begin{equation}
\setcounter{equation}{10}
\begin{split}{}
&{\cal L}\left( {{{\left\{ {{{\bf{x}}^t}} \right\}}_{t \in {\cal T}}},r ,\left\{ {{\mu _k}} \right\}_{k = 1}^K} \right)\\
 =& r  + \sum\nolimits_{k = 1}^K {{\mu _k}\left( {\overline {{r_k}}  - r } \right)} = \left( {1 - \sum\nolimits_{k = 1}^K {{\mu _k}} } \right)r  \\
 &+ \frac{1}{T}\int_0^T {\sum\nolimits_{k = 1}^K {{\mu _k}{{\log }_2}\left( {1 + {\gamma _k}({{\bf{x}}^t})} \right)dt} }.
\end{split}
\end{equation}

Armed with (10), the dual function of (P2.1) is obtained as
\begin{equation}
\begin{split}{}
&f(\left\{ {{\mu _k}} \right\}_{k = 1}^K) \\
=& \mathop {\max }\limits_{{{\left\{ {{{\bf{x}}^t}} \right\}}_{t \in {\cal T}}},r } {\cal L}\left( {{{\left\{ {{{\bf{x}}^t}} \right\}}_{t \in {\cal T}}},r ,\left\{ {{\mu _k}} \right\}_{k = 1}^K} \right), {\rm{s}}.{\rm{t}}., (7{\rm{c}}), (7{\rm{d}}).
\end{split}
\end{equation}

Observing (10) and (11), to ensure that $f(\left\{ {{\mu _k}} \right\}_{k = 1}^K)$ has the finite upper bound, it must have $\sum\nolimits_{k = 1}^K {{\mu _k}}  = 1$. Otherwise if $\sum\nolimits_{k = 1}^K {{\mu _k}}  > 1$ or $\sum\nolimits_{k = 1}^K {{\mu _k}}  < 1$, $f(\left\{ {{\mu _k}} \right\}_{k = 1}^K)$ would tend to infinity by setting $r \to \infty $ or $r \to -\infty $.

Based on the above analysis, the dual problem of (P2.1) can be expressed as
 \begin{align}
&({\rm{D2.1}}):{\rm{  }}\mathop {\min }\limits_{{\left\{ {{\mu _k}} \right\}_{k = 1}^K}} \ f(\left\{ {{\mu _k}} \right\}_{k = 1}^K)  \tag{${\rm{12a}}$}\\
{\rm{              }}&{\rm{s.t.}} \ \ \ \ \sum\nolimits_{k = 1}^K {{\mu _k}}  = 1,\tag{${\rm{12b}}$}\\
 &\ \ \ \ \ \ \ \ {\mu _k} > 0,k \in {\cal K}.\tag{${\rm{12c}}$}
\end{align}

After obtaining the structure of the dual problem, problem (P2.1) now can be solved by executing two steps alternately until ${\left\{ {{\mu _k}} \right\}_{k = 1}^K}$ converge within a prescribed tolerance. The first step is to solve the problem in (11) for obtaining $f(\left\{ {{\mu _k}} \right\}_{k = 1}^K)$ under any given and feasible dual variables ${\left\{ {{\mu _k}} \right\}_{k = 1}^K}$. The second step is to solve (D2.1) to find the optimal ${\left\{ {{\mu _k}} \right\}_{k = 1}^K}$ for minimizing $f(\left\{ {{\mu _k}} \right\}_{k = 1}^K)$. Finally, the solutions to (P2.1) can be constructed based on the obtained results in the above.

\textbf{1) The step 1:} For any feasible ${\left\{ {{\mu _k}} \right\}_{k = 1}^K}$ satisfying $\sum\nolimits_{k = 1}^K {{\mu _k}}  = 1$, note that $\left( {1 - \sum\nolimits_{k = 1}^K {{\mu _k}} } \right)r  = 0$ regardless of how $r$ is. Based on this fact, the problem in (11) can be simplified as
\begin{equation}
\setcounter{equation}{13}
\begin{split}{}
&f(\left\{ {{\mu _k}} \right\}_{k = 1}^K) = \mathop {\max }\limits_{{{\left\{ {{{\bf{x}}^t}} \right\}}_{t \in {\cal T}}}} \frac{1}{T}\\
&\times \int_0^T {\sum\nolimits_{k = 1}^K {{\mu _k}{{\log }_2}\left( {1 + {\gamma _k}({{\bf{x}}^t})} \right)dt} } , \ {\rm{s}}.{\rm{t}}., (7{\rm{c}}), (7{\rm{d}}),
\end{split}
\end{equation}
which can be further equivalently simplified as the following problem by dropping the time index, i.e.,
\begin{equation}
\begin{split}{}
&f(\left\{ {{\mu _k}} \right\}_{k = 1}^K) \\
=& \mathop {\max }\limits_{\bf{x}} \ \sum\nolimits_{k = 1}^K {{\mu _k}{{\log }_2}\left( {1 + {\gamma _k}({\bf{x}})} \right)}   \\
&{\rm{                 }}\ {\rm{s}}.{\rm{t}}. \ {\left[ {\bf{x}} \right]_m} - {\left[ {\bf{x}} \right]_p} \ge {d_{\min }},\forall m,p \in {\cal M},m > p,\\
&{\rm{                     }}\ \ \ \ \ \ {\left[ {\bf{x}} \right]_m} \in [0,L],\forall m \in {\cal M}.
\end{split}
\end{equation}

The problem in (14) is highly non-convex due to the complex structure of ${{\gamma _k}({\bf{x}})}$, $\forall k \in {\cal K}$. In the next, the SCA technique will be exploited for solving this problem. To proceed, one important Lemma obtained from our previous work \cite{SZH_TVT} is first provided as follows.

\textbf{Lemma 1 \cite{SZH_TVT}}: For any $k$, ${\widetilde \gamma _k}({\bf{x}}) \buildrel \Delta \over = {\bf{h}}_k^H({\bf{x}}){\bf{A}}_k^{ - 1}({\bf{x}}){{\bf{h}}_k}({\bf{x}})$ can be equivalently expressed as
\begin{equation}
\begin{split}{}
{\widetilde \gamma _k}({\bf{x}}) \buildrel \Delta \over = \mathop {\max }\limits_{{\bf{z}} \in {{\mathbb{C}}^{MN \times 1}}} \left\{ \begin{array}{l}
2{\mathop{\rm Re}\nolimits} \left( {{\bf{h}}_k^H({\bf{x}}){\bf{z}}} \right)\\
 - {{\bf{z}}^H}{{\bf{A}}_k}({\bf{x}}){\bf{z}}
\end{array} \right\},
\end{split}
\end{equation}
where ${\bf{z}}$ is an auxiliary variable and the optimal ${\bf{z}}$ is ${{\bf{z}}^*} = {\bf{A}}_k^{ - 1}({\bf{x}}){{\bf{h}}_k}({\bf{x}})$.

Based on Lemma 1, the lower bound of ${\widetilde \gamma _k}({\bf{x}})$ with the given ${{\bf{x}}_{(q)}}$ can be derived as
\begin{equation}
\begin{split}{}
&{\widetilde \gamma _k}({\bf{x}})) \ge {\widetilde \gamma _k}({\bf{x}},{{\bf{x}}_{(q)}})\\
 =& 2{\mathop{\rm Re}\nolimits} \left( {{\bf{h}}_k^H({\bf{x}}){\bf{A}}_k^{ - 1}({{\bf{x}}_{(q)}}){{\bf{h}}_k}({{\bf{x}}_{(q)}})} \right)\\
 -& {\left( {{\bf{A}}_k^{ - 1}({{\bf{x}}_{(q)}}){{\bf{h}}_k}({{\bf{x}}_{(q)}})} \right)^H}{{\bf{A}}_k}({\bf{x}}){\bf{A}}_k^{ - 1}({{\bf{x}}_{(q)}}){{\bf{h}}_k}({{\bf{x}}_{(q)}}),
\end{split}
\end{equation}
where the equality is established only when ${{\bf{x}}_{(q)}} = {\bf{x}}$.

  \begin{figure*}[b!]
  \hrulefill
\setcounter{equation}{27}
\begin{equation}
\begin{split}{}
{\widetilde \gamma _k}({\bf{x}})) \ge & {\widetilde \gamma _k}^\prime ({\bf{x}},{{\bf{x}}_{(q)}}) = 2\sqrt {{\beta _k}} \sum\nolimits_{m = 1}^M {\sum\nolimits_{n = 1}^N {\left| {{{\left[ {{{\bf{g}}_k}({{\bf{x}}_{(q)}})} \right]}_{m,n}}} \right|{f_{1,m,n,k}}} } \\
& + 2\sum\nolimits_{j = 1,j \ne k}^K {\overline {{P_j}} \sqrt {{\beta _j}} \sum\nolimits_{m = 1}^M {\sum\nolimits_{n = 1}^N {\left| {{{\left[ {{{\bf{o}}_j}({{\bf{x}}_{(q)}}} \right]}_{m,n}}} \right|{f_{2,m,n,j}}} } }  - {D_k}.
\end{split}
\end{equation}
\end{figure*}

Subsequently, by denoting ${\bf{g}}({{\bf{x}}_{(q)}}) \buildrel \Delta \over = {\bf{A}}_k^{ - 1}({{\bf{x}}_{(q)}}){{\bf{h}}_k}({{\bf{x}}_{(q)}})$ and expanding ${{\bf{A}}_k}({\bf{x}})$, we can re-express ${\widetilde \gamma _k}({\bf{x}},{{\bf{x}}_{(q)}})$ as
\begin{equation}
\setcounter{equation}{17}
\begin{split}{}
&{\widetilde \gamma _k}({\bf{x}},{{\bf{x}}_{(q)}})\\
 =& 2{\mathop{\rm Re}\nolimits} \left( {{\bf{h}}_k^H({\bf{x}}){{\bf{g}}_k}({{\bf{x}}_{(q)}})} \right) - {\bf{g}}_k^H({{\bf{x}}_{(q)}}){{\bf{A}}_k}({\bf{x}}){{\bf{g}}_k}({{\bf{x}}_{(q)}})\\
 =& 2{\mathop{\rm Re}\nolimits} \left( {{\bf{h}}_k^H({\bf{x}}){{\bf{g}}_k}({{\bf{x}}_{(q)}})} \right) - {\bf{g}}_k^H({{\bf{x}}_{(q)}})\\
 &\times \left( {\sum\nolimits_{j = 1,j \ne k}^K {\overline {{P_j}} } {{\bf{h}}_j}({\bf{x}}){\bf{h}}_j^H({\bf{x}}) + {{\bf{I}}_{MN}}} \right){{\bf{g}}_k}({{\bf{x}}_{(q)}})\\
 =& 2{\mathop{\rm Re}\nolimits} \left( {{\bf{h}}_k^H({\bf{x}}){{\bf{g}}_k}({{\bf{x}}_{(q)}})} \right) - {\left\| {{{\bf{g}}_k}({{\bf{x}}_{(q)}})} \right\|^2}\\
 &- \sum\nolimits_{j = 1,j \ne k}^K {\overline {{P_j}} } {\bf{h}}_j^H({\bf{x}})\underbrace {{{\bf{g}}_k}({{\bf{x}}_{(q)}}){\bf{g}}_k^H({{\bf{x}}_{(q)}})}_{{{\bf{G}}_k}({{\bf{x}}_{(q)}})}{{\bf{h}}_j}({\bf{x}}),
\end{split}
\end{equation}
where each term ${\bf{h}}_j^H({\bf{x}}){{\bf{G}}_k}({{\bf{x}}_{(q)}}){{\bf{h}}_j}({\bf{x}})$ ($j = 1,...,K,j \ne k$) shown in (17) can be upper-bounded by \cite{MA_MIMO1}
\begin{equation}
\begin{split}{}
&{\bf{h}}_j^H({\bf{x}}){{\bf{G}}_k}({{\bf{x}}_{(q)}}){{\bf{h}}_j}({\bf{x}}) \le {\bf{h}}_j^H({\bf{x}}){{\bf{\Theta }}_k}({{\bf{x}}_{(q)}}){{\bf{h}}_j}({\bf{x}})\\
 -& 2{\mathop{\rm Re}\nolimits} \left\{ {{\bf{h}}_j^H({\bf{x}})\left( {{{\bf{\Theta }}_k}({{\bf{x}}_{(q)}}) - {{\bf{G}}_k}({{\bf{x}}_{(q)}})} \right){{\bf{h}}_j}({{\bf{x}}_{(q)}})} \right\}\\
 +& {\bf{h}}_j^H({{\bf{x}}_{(q)}})\left( {{{\bf{\Theta }}_k}({{\bf{x}}_{(q)}}) - {{\bf{G}}_k}({{\bf{x}}_{(q)}})} \right){{\bf{h}}_j}({{\bf{x}}_{(q)}}),
\end{split}
\end{equation}
with ${{\bf{\Theta }}_k}({{\bf{x}}_{(q)}}) = \lambda _{k,\max }^{(q)}{{\bf{I}}_{MN}}$ and $\lambda _{k,\max }^{(q)}$ being the maximum eigenvalue of ${{{\bf{G}}_k}({{\bf{x}}_{(q)}})}$. In addition, ${\bf{h}}_j^H({\bf{x}}){{\bf{\Theta }}_k}({{\bf{x}}_{(q)}}){{\bf{h}}_j}({\bf{x}})$ in (18) can be simplified as
\begin{equation}
\begin{split}{}
&{\bf{h}}_j^H({\bf{x}}){{\bf{\Theta }}_k}({{\bf{x}}_{(q)}}){{\bf{h}}_j}({\bf{x}}) = \lambda _{k,\max }^{(q)}{\left\| {{{\bf{h}}_j}({\bf{x}})} \right\|^2}\\
 =& \lambda _{k,\max }^{(q)}{\beta _j}{\left\| {{{\left( {{\bf{f}}_j^{{\rm{hor}}}({\bf{x}})} \right)}^H} \odot {{\left( {{\bf{f}}_j^{{\rm{ver}}}({{\bf{y}}^{{\rm{FPA}}}})} \right)}^H}} \right\|^2}\\
 =& \lambda _{k,\max }^{(q)}{\beta _j}{\left\| {{{\left( {{\bf{f}}_j^{{\rm{ver}}}({{\bf{y}}^{{\rm{FPA}}}})} \right)}^H} \otimes {{\left( {{\bf{f}}_j^{{\rm{hor}}}({\bf{x}})} \right)}^H}} \right\|^2}\\
\mathop  = \limits^{(a)} & \lambda _{k,\max }^{(q)}{\beta _j}{\left\| {{\bf{f}}_j^{{\rm{ver}}}({{\bf{y}}^{{\rm{FPA}}}})} \right\|^2}{\left\| {{{\left( {{\bf{f}}_j^{{\rm{hor}}}({\bf{x}})} \right)}^H}} \right\|^2}\\
 =& \lambda _{k,\max }^{(q)}{\beta _j}MN,
\end{split}
\end{equation}
where the equality (a) is established since ${\left\| {{\bf{a}} \otimes {\bf{b}}} \right\|^2} = {\left\| {\bf{a}} \right\|^2}{\left\| {\bf{b}} \right\|^2}$.

Based on (18) and (19), we can derive the upper bound of ${\bf{h}}_j^H({\bf{x}}){{\bf{G}}_k}({{\bf{x}}_{(q)}}){{\bf{h}}_j}({\bf{x}})$ as \begin{equation}
\begin{split}{}
&{\bf{h}}_j^H({\bf{x}}){{\bf{G}}_k}({{\bf{x}}_{(q)}}){{\bf{h}}_j}({\bf{x}}) \\ \le &  - 2{\mathop{\rm Re}\nolimits} \left\{ {{\bf{h}}_j^H({\bf{x}}){{\bf{o}}_j}({{\bf{x}}_{(q)}})} \right\} + {C_j},
\end{split}
\end{equation}
where ${{\bf{o}}_j}({{\bf{x}}_{(q)}}) = \left( {{{\bf{\Theta }}_k}({{\bf{x}}_{(q)}}) - {{\bf{G}}_k}({{\bf{x}}_{(q)}})} \right){{\bf{h}}_j}({{\bf{x}}_{(q)}})$ and ${C_j} = \lambda _{k,\max }^{(q)}{\beta _j}MN + {\bf{h}}_j^H({{\bf{x}}_{(q)}})\left( {{{\bf{\Theta }}_k}({{\bf{x}}_{(q)}}) - {{\bf{G}}_k}({{\bf{x}}_{(q)}})} \right){{\bf{h}}_j}({{\bf{x}}_{(q)}})$.

Based on (17) and (20), we can obtain
\begin{equation}
\begin{split}{}
&{\widetilde \gamma _k}({\bf{x}},{{\bf{x}}_{(q)}})\\
 \ge & 2{\mathop{\rm Re}\nolimits} \left( {{\bf{h}}_k^H({\bf{x}}){{\bf{g}}_k}({{\bf{x}}_{(q)}})} \right) - {\left\| {{{\bf{g}}_k}({{\bf{x}}_{(q)}})} \right\|^2}\\
 & - \sum\nolimits_{j = 1,j \ne k}^K {\overline {{P_j}} } \left( { - 2{\mathop{\rm Re}\nolimits} \left\{ {{\bf{h}}_j^H({\bf{x}}){{\bf{o}}_j}({{\bf{x}}_{(q)}})} \right\} + {C_j}} \right)\\
 =& 2{\mathop{\rm Re}\nolimits} \left( {{\bf{h}}_k^H({\bf{x}}){{\bf{g}}_k}({{\bf{x}}_{(q)}})} \right)  \\
 &+ 2\sum\nolimits_{j = 1,j \ne k}^K {\overline {{P_j}} {\mathop{\rm Re}\nolimits} \left\{ {{\bf{h}}_j^H({\bf{x}}){{\bf{o}}_j}({{\bf{x}}_{(q)}})} \right\}}  - {D_k},
\end{split}
\end{equation}
where ${D_k} = \sum\nolimits_{j = 1,j \ne k}^K {\overline {{P_j}} } {C_j} + {\left\| {{{\bf{g}}_k}({{\bf{x}}_{(q)}})} \right\|^2}$.

Since the lower bound derived in (21) remains non-convex w.r.t. ${\bf{x}}$, we require further transformations as described below.

Specifically, based on the structure of ${{\bf{h}}_k}({\bf{x}})$ presented in (1), we can obtain the $i$-th element in ${{\bf{h}}_k}({\bf{x}})$ as
\begin{equation}
\begin{split}{}
{\left[ {{{\bf{h}}_k}({\bf{x}})} \right]_i} = \sqrt {{\beta _k}} {e^{ - j\frac{{2\pi }}{\lambda }\left( {{{\left[ {\bf{x}} \right]}_m}{\vartheta _k} + {{\left[ {\bf{y}} \right]}_n}{\varphi _k}} \right)}},
\end{split}
\end{equation}
where $m = \left\lfloor {(i - 1)/N} \right\rfloor  + 1$ and $n = i - (m - 1)N$. Based on (22), we can expand ${\mathop{\rm Re}\nolimits} \left( {{\bf{h}}_k^H({\bf{x}}){{\bf{g}}_k}({{\bf{x}}_{(q)}})} \right)$ and each ${{\mathop{\rm Re}\nolimits} \left\{ {{\bf{h}}_j^H({\bf{x}}){{\bf{o}}_j}({{\bf{x}}_{(q)}})} \right\}}$ ($j = 1,...,K,j \ne k$) in (21) as
\begin{equation} \small
\begin{split}{}
&{\mathop{\rm Re}\nolimits} \left( {{\bf{h}}_k^H({\bf{x}}){{\bf{g}}_k}({{\bf{x}}_{(q)}})} \right) = \sqrt {{\beta _k}} \\
 \times & \sum\nolimits_{m = 1}^M {\sum\nolimits_{n = 1}^N {\left| {{{\left[ {{{\bf{g}}_k}({{\bf{x}}_{(q)}})} \right]}_{m,n}}} \right|\cos \left( {{\zeta _{m,n,k}} - {\psi _{m,n,k}}} \right)} } ,
\end{split}
\end{equation}
\begin{equation} \small
\begin{split}{}
&{\mathop{\rm Re}\nolimits} \left( {{\bf{h}}_j^H({\bf{x}}){{\bf{o}}_j}({{\bf{x}}_{(q)}})} \right) = \sqrt {{\beta _j}} \\
\times & \sum\nolimits_{m = 1}^M {\sum\nolimits_{n = 1}^N {\left| {{{\left[ {{{\bf{o}}_j}({{\bf{x}}_{(q)}}} \right]}_{m,n}}} \right|\cos \left( {{\zeta _{m,n,j}} - {\psi _{m,n,j}}^\prime } \right)} } ,
\end{split}
\end{equation}
where
\begin{equation} \nonumber
\begin{split}{}
{\left[ {{{\bf{g}}_k}({{\bf{x}}_{(q)}})} \right]_{m,n}} \buildrel \Delta \over =& {\left[ {{{\bf{g}}_k}({{\bf{x}}_{(q)}})} \right]_{(m - 1)N + n}},\\
{\left[ {{{\bf{o}}_j}({{\bf{x}}_{(q)}}} \right]_{m,n}} \buildrel \Delta \over =& {\left[ {{{\bf{o}}_j}({{\bf{x}}_{(q)}}} \right]_{(m - 1)N + n}},\\
{\zeta _{m,n,k}} =& \frac{{2\pi }}{\lambda }\left( {{{\left[ {\bf{x}} \right]}_m}{\vartheta _k} + {{\left[ {\bf{y}} \right]}_n}{\varphi _k}} \right),\\
{\psi _{m,n,k}} =& \arg \left( {{{\left[ {{{\bf{g}}_k}({{\bf{x}}_{(q)}})} \right]}_{m,n}}} \right),\\
{\psi _{m,n,j}}^\prime  =& \arg \left( {{{\left[ {{{\bf{o}}_j}({{\bf{x}}_{(q)}}} \right]}_{m,n}}} \right).
\end{split}
\end{equation}

Then, based on the second-order Taylor expansion, we can obtain that with the given ${x_{(q)}}$, it has
\begin{equation}
\begin{split}{}
&\cos (ax - b) \ge \cos (a{x_{(q)}} - b)\\
 -& a\sin (a{x_{(q)}} - b)(x - {x_{(q)}}) - \frac{{{a^2}}}{2}{(x - {x_{(q)}})^2},
\end{split}
\end{equation}
where $a > 0$ and $b$ are constants. Armed with (25), the lower bounds of ${\mathop{\rm Re}\nolimits} \left( {{\bf{h}}_k^H({\bf{x}}){{\bf{g}}_k}({{\bf{x}}_{(q)}})} \right)$ and ${\mathop{\rm Re}\nolimits} \left( {{\bf{h}}_j^H({\bf{x}}){{\bf{o}}_j}({{\bf{x}}_{(q)}})} \right)$ shown in (23) and (24) can be derived as
\begin{equation}
\begin{split}{}
&{\mathop{\rm Re}\nolimits} \left( {{\bf{h}}_k^H({\bf{x}}){{\bf{g}}_k}({{\bf{x}}_{(q)}})} \right) \ge \sqrt {{\beta _k}} \\
 \times &\sum\nolimits_{m = 1}^M {\sum\nolimits_{n = 1}^N {\left| {{{\left[ {{{\bf{g}}_k}({{\bf{x}}_{(q)}})} \right]}_{m,n}}} \right|{f_{1,m,n,k}}} },
\end{split}
\end{equation}
\begin{equation}
\begin{split}{}
&{\mathop{\rm Re}\nolimits} \left( {{\bf{h}}_j^H({\bf{x}}){{\bf{o}}_j}({{\bf{x}}_{(q)}})} \right) \ge \sqrt {{\beta _j}} \\
 \times & \sum\nolimits_{m = 1}^M {\sum\nolimits_{n = 1}^N {\left| {{{\left[ {{{\bf{o}}_j}({{\bf{x}}_{(q)}}} \right]}_{m,n}}} \right|{f_{2,m,n,j}}} },
\end{split}
\end{equation}
where
\begin{equation} \nonumber
\begin{split}{}
{f_{1,m,n,k}} =& \cos \left( {{\zeta _{m,n,k,(q)}} - {\psi _{m,n,k}}} \right)\\
 &- \frac{{2\pi }}{\lambda }{\vartheta _k}\sin \left( {{\zeta _{m,n,k,(q)}} - {\psi _{m,n,k}}} \right)\\
 &\times \left( {{{\left[ {\bf{x}} \right]}_m} - {{\left[ {{{\bf{x}}_{(q)}}} \right]}_m}} \right) - \frac{{{\delta _k}}}{2}{\left( {{{\left[ {\bf{x}} \right]}_m} - {{\left[ {{{\bf{x}}_{(q)}}} \right]}_m}} \right)^2},
\end{split}
\end{equation}
\begin{equation} \nonumber
\begin{split}{}
{f_{2,m,n,j}} =& \cos \left( {{\zeta _{m,n,j,(q)}} - {\psi _{m,n,j}}^\prime } \right)\\
 &- \frac{{2\pi }}{\lambda }{\vartheta _j}\sin \left( {{\zeta _{m,n,j,(q)}} - {\psi _{m,n,j}}^\prime } \right)\\
 &\times \left( {{{\left[ {\bf{x}} \right]}_m} - {{\left[ {{{\bf{x}}_{(q)}}} \right]}_m}} \right) - \frac{{{\delta _j}}}{2}{\left( {{{\left[ {\bf{x}} \right]}_m} - {{\left[ {{{\bf{x}}_{(q)}}} \right]}_m}} \right)^2},
\end{split}
\end{equation}
with ${\zeta _{m,n,k,(q)}} = \frac{{2\pi }}{\lambda }\left( {{{\left[ {{{\bf{x}}_{(q)}}} \right]}_m}{\vartheta _k} + {{\left[ {\bf{y}} \right]}_n}{\varphi _k}} \right)$ and ${\delta _k} = {\left( {\frac{{2\pi }}{\lambda }{\vartheta _k}} \right)^2}$.

Based on (21), (26) and (27), with the given ${{{\bf{x}}_{(q)}}}$, we can finally obtain the convex lower bound of ${\widetilde \gamma _k}({\bf{x}})$, i,e., ${\widetilde \gamma _k}^\prime ({\bf{x}},{{\bf{x}}_{(q)}})$, as in (28).

With ${\widetilde \gamma _k}^\prime ({\bf{x}},{{\bf{x}}_{(q)}})$ at hand, the problem in (14) can be relaxed as the following convex problem,
\begin{equation}
\setcounter{equation}{29}
\begin{split}{}
&\mathop {\max }\limits_{\bf{x}} \;\sum\nolimits_{k = 1}^K {{\mu _k}{{\log }_2}\left( {1 + \overline {{P_k}} {{\widetilde \gamma }_k}^\prime ({\bf{x}},{{\bf{x}}_{(q)}})} \right)} \\
&\ {\rm{s}}.{\rm{t}}.\;{\left[ {\bf{x}} \right]_m} - {\left[ {\bf{x}} \right]_p} \ge {d_{\min }},\forall m,p \in {\cal M},m > p,\\
 \ & \quad \ \;\;{\left[ {\bf{x}} \right]_m} \in [0,L],\forall m \in {\cal M},
\end{split}
\end{equation}
which can be solved efficiently with the CVX tool.

The problem in (29) can be solved iteratively until the objective converges to a stationary value. Note that a different initial deployment pattern ${{{\bf{x}}_{(1)}}}$ may lead to a different stationary output, denoted as ${{\bf{x}}_{{\rm{sta}}}}$. Therefore, a common and practical strategy is to run the iteration from $S$ distinct, randomly chosen initial points, denoted as $\left\{ {{{\bf{x}}_{(1,1)}},...,{{\bf{x}}_{(1,S)}}} \right\}$. The corresponding stationary solutions are collected into a set $\Omega  = \left\{ {{{\bf{x}}_{({\rm{sta}},1)}},...,{{\bf{x}}_{({\rm{sta}},S)}}} \right\}$. Afterwards, from $\Omega$, those solutions that can concurrently maximize the objective in (14) are selected and placed into a new set ${{\Omega _{\left\{ {{\mu _k}} \right\}_{k = 1}^K}}}$, where we denote ${\left[ {{\Omega _{\left\{ {{\mu _k}} \right\}_{k = 1}^K}}} \right]_i} = {{\bf{x}}_{\left\{ {{\mu _k}} \right\}_{k = 1}^K,i}}$ and $\left| {{\Omega _{\left\{ {{\mu _k}} \right\}_{k = 1}^K}}} \right| = \Gamma $.

 Accordingly, given ${\left\{ {{\mu _k}} \right\}_{k = 1}^K}$, the solution to the problem in (14) can be obtained as
 \begin{equation}
\begin{split}{}
{\bf{x}}_{\left\{ {{\mu _k}} \right\}_{k = 1}^K}^* = {{\bf{x}}_{\left\{ {{\mu _k}} \right\}_{k = 1}^K,i}},\forall i = 1,..,\Gamma .
\end{split}
\end{equation}

By substituting ${\bf{x}}_{\left\{ {{\mu _k}} \right\}_{k = 1}^K}^*$ into the problem in (14), the dual function $f(\left\{ {{\mu _k}} \right\}_{k = 1}^K)$ can be obtained.

\textbf{2) The step 2:} With $f(\left\{ {{\mu _k}} \right\}_{k = 1}^K)$ at hand, we need to solve the dual problem (D2.1) for determining the optimal $\left\{ {{\mu _k}} \right\}_{k = 1}^K$ to minimize $f(\left\{ {{\mu _k}} \right\}_{k = 1}^K)$.

Specifically, since the dual function $f(\left\{ {{\mu _k}} \right\}_{k = 1}^K)$ is always convex but generally non-differentiable, problem (D2.1) can be solved by subgradient-based methods such as the ellipsoid method \cite{Xujie_UAV}. The details are provided as follows.

First, the subgradient of the objective should be computed by discussing two different cases:
\begin{itemize}
\item[$\bullet$]  Case 1: There exist the negative value(s) in the current dual variables $\left\{ {{\mu _k}} \right\}_{k = 1}^K$. Denote $m$ as the index of the smallest element in $\left\{ {{\mu _k}} \right\}_{k = 1}^K$. Then, the subgradient in this case should be
     \begin{equation}
\begin{split}{}
{\bf{gra}} =  - {{\bf{e}}_m},
\end{split}
\end{equation}
where the $m$-th element in ${{\bf{e}}_m} \in {{\mathbb{R}}^{K \times 1}}$ is 1 and the other elements are 0.

\item[$\bullet$]  Case 2: The current dual variables $\left\{ {{\mu _k}} \right\}_{k = 1}^K > 0$. If $\sum\nolimits_{k = 1}^K {{\mu _k}}  > 1$, the subgradient should be
         \begin{equation}
\begin{split}{}
{\bf{gra}} =  {\bf{e}},
\end{split}
\end{equation}
where the elements in ${{\bf{e}}} \in {{\mathbb{R}}^{K \times 1}}$ are all set as 1. If $\sum\nolimits_{k = 1}^K {{\mu _k}}  <1$, the subgradient should be
  \begin{equation}
\begin{split}{}
{\bf{gra}} = - {\bf{e}}.
\end{split}
\end{equation}

Otherwise if $\sum\nolimits_{k = 1}^K {{\mu _k}}  =1$, the subgradient should be
  \begin{equation} \small
\begin{split}{}
&{\bf{gra}} \\
=& {\left[ {{{\log }_2}(1 + {\gamma _1}({\bf{x}}_{\left\{ {{\mu _k}} \right\}_{k = 1}^K}^*)),...,{{\log }_2}(1 + {\gamma _K}({\bf{x}}_{\left\{ {{\mu _k}} \right\}_{k = 1}^K}^*))} \right]^T}.
\end{split}
\end{equation}
\end{itemize}

With the subgradient at hand, the following step is to update the dual variables and the positive definite matrix (denoted as ${\bf{B}}$, which is initially set as ${\bf{B}} = K \times {\rm{diag}}\left\{ {\bf{e}} \right\}$) as \cite{Xujie_UAV}
  \begin{equation}
\begin{split}{}
{\mu _k} =& {\mu _k} - \frac{1}{{K + 1}}{\left[ {\frac{{{\bf{Bgra}}}}{{\sqrt {{\bf{gr}}{{\bf{a}}^T}{\bf{Bgra}}} }}} \right]_k},k \in {\cal K},\\
{\bf{B}} =& \frac{{{K^2}}}{{{K^2} - 1}}\left( {{\bf{B}} - \frac{2}{{K + 1}}\frac{{{\bf{Bgra}} \times {\bf{gr}}{{\bf{a}}^T}{\bf{B}}}}{{{\bf{gr}}{{\bf{a}}^T}{\bf{Bgra}}}}} \right).
\end{split}
\end{equation}

\begin{algorithm}
\caption{The Procedures for repeating steps 1 and 2.}
  \begin{algorithmic}[1]

\State \textbf{{Initialization:}} ${\mu _k} = 1/K$, $k = 1,...,K$, and the positive definite matrix ${\bf{B}} = K \times {\rm{diag}}\left\{ {\bf{e}} \right\}$.

\State \textbf{Repeat:}

\State \quad \textbf{Step 1}: Solve the problem in (29) iteratively from $S$ initial points $\left\{ {{{\bf{x}}_{(1,1)}},...,{{\bf{x}}_{(1,S)}}} \right\}$ and output ${\bf{x}}_{\left\{ {{\mu _k}} \right\}_{k = 1}^K}^* = {{\bf{x}}_{\left\{ {{\mu _k}} \right\}_{k = 1}^K,i}},\forall i = 1,..,\Gamma .$

\State \quad \textbf{Step 2}: Determine the subgradient ${\bf{gra}}$ via (31)-(34). Then update $\left\{ {{\mu _k}} \right\}_{k = 1}^K$ and the positive definite matrix ${\bf{B}}$ via (35).

\State \textbf{Until:} $\left\{ {{\mu _k}} \right\}_{k = 1}^K$ converge with a prescribed accuracy.
  \end{algorithmic}
\end{algorithm}

Based on the above analysis, we summary the procedures for alternatively implementing steps 1 and 2 as presented in Algorithm 1, which finally outputs the stationary $\left\{ {{\mu _k}} \right\}_{k = 1}^K$ and ${{\bf{x}}_{\left\{ {{\mu _k}} \right\}_{k = 1}^K,i}}$, $i = 1,..,\Gamma $.

\textbf{3) Construct the solutions to (P2.1)}: With $\Gamma $ antenna deployment patterns obtained from Algorithm 1, the BS needs to time-share among these patterns to determine the solutions to (P2.1). Here, the time-sharing implies that the BS should adopt each deployment pattern for a certain fraction of the total period $T$. Denoting by ${t_i}$ the deployment duration allocated to the pattern ${{\bf{x}}_{\left\{ {{\mu _k}} \right\}_{k = 1}^K,i}}$, the optimal time allocations can be obtained by solving the following problem.
 \begin{align}
&({\rm{P3}}):{\rm{  }}\mathop {\max }\limits_{\left\{ {{t_i}} \right\}_{i = 1}^\Gamma , r } \ r  \tag{${\rm{36a}}$}\\
{\rm{              }}&{\rm{s.t.}} \ \ \ \ \frac{1}{T}\sum\nolimits_{i = 1}^\Gamma  {{t_i}{{\log }_2}(1 + {\gamma _k}({{\bf{x}}_{\left\{ {{\mu _k}} \right\}_{k = 1}^K,i}}))}  \ge r ,\forall k \in {\cal K},\tag{${\rm{36b}}$}\\
 &\ \ \ \ \ \ \ \sum\nolimits_{i = 1}^\Gamma  {{t_i}}  = T,\tag{${\rm{36c}}$}
\end{align}
which is a simple linear program problem and can be solved with the CVX tool.

\textbf{Complexity Analysis:} For step 1, there are $M$ real variables for the problem in (29). Therefore, the complexity of solving it from $S$ initial points is about ${\cal O}({I_{{\rm{inner}}}}S{M^{3.5}})$, where ${I_{{\rm{inner}}}}$ denotes the number of iterations. For step 2, the complexity of updating $\left\{ {{\mu _k}} \right\}_{k = 1}^K$ and ${\bf{B}}$ in (35) is easily summarized as ${\cal O}({K^2})$. Therefore, the procedures of repeating steps 1 and 2 own the complexity of ${\cal O}({I_{{\rm{outer}}}}({I_{{\rm{inner}}}}S{M^{3.5}} + {K^2}))$, where ${I_{{\rm{outer}}}}$ is the number of outer iterations. Similarly, the complexity of solving (P3) is about ${\cal O}({(\Gamma  + 1)^{3.5}})$. Based on these facts, the total complexity of solving (P2.1) is about ${\cal O}({I_{{\rm{outer}}}}({I_{{\rm{inner}}}}S{M^{3.5}} + {K^2}) + {(\Gamma  + 1)^{3.5}})$.

\textbf{Remark 1:} From the above analysis, in the ideal case where the movement speed of each sliding track is unlimited, the BS does not need to switch its antenna deployment pattern frequently. Instead, it only needs to adopt a finite set of deployment patterns to achieve the optimal performance. \textbf{This observation perfectly fits the inherent characteristics of MA systems, as infrequent antenna movements will greatly reduce the associated power consumption.} Therefore, the proposed deployment strategy not only provides a tractable solution, but also offers a practical and energy-efficient approach for MAs-enabled communication systems.

 \begin{figure} [!t]
\centering
\includegraphics[width=8cm]{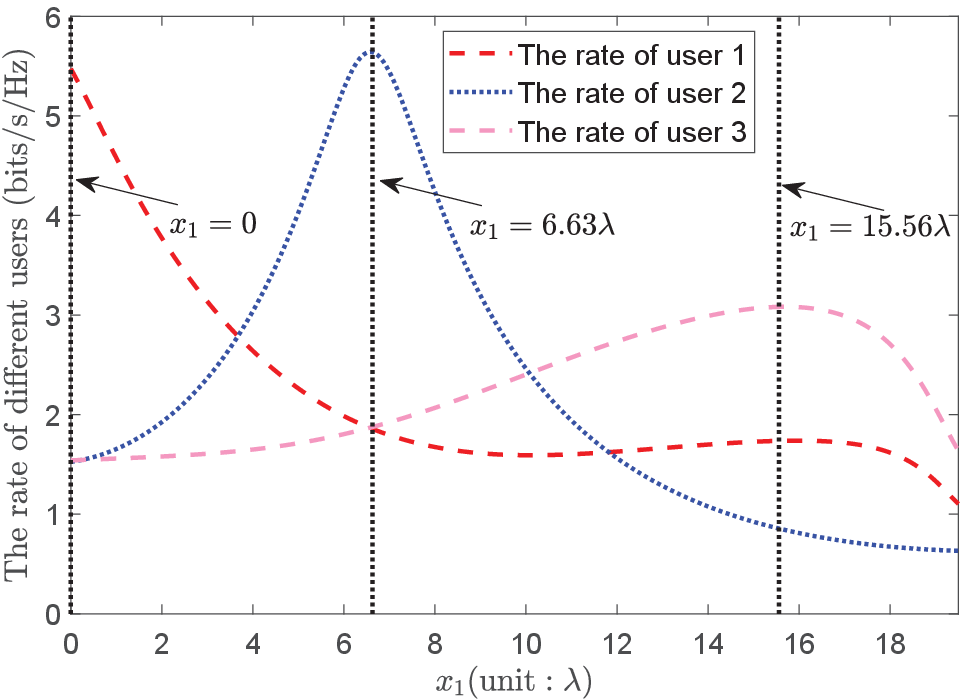}
\captionsetup{font=small}
\caption{The achievable rate of different users w.r.t. $x_1$.} \label{fig:Fig1}
\end{figure}

\subsection{The Special Case of $M = 2$}
In this subsection, we aim to provide certain insights for the conclusions obtained above, by considering the special case of $M = 2$ vertical sliding tracks. In this case, we can conveniently fix the horizontal coordinate of the second track as $x_2 = L$.\footnote{This is because each ${\gamma _k}({\bf{x}} = [{x_1},{x_2}])$, $k \in {\cal K}$, depends only on the difference between $x_2$ and $x_1$. Therefore, setting $x_2$ to its maximum value provides $x_1$ with the greatest flexibility for optimization.} Then, each different deployment pattern requiring time-sharing can be just reflected by the corresponding horizontal coordinate ($x_1$) of the first track. For satisfying the constraints in (7c) and (7d), $x_1$ should lie in the interval $[0:L - {d_{\min }}]$.

When $M = 2$, it is noted that the step 1 in Algorithm 1 given $\left\{ {{\mu _k}} \right\}_{k = 1}^K$ can be simplified to find the optimal $x_1$ that maximizes $\sum\nolimits_{k = 1}^K {{\mu _k}{{\log }_2}\left( {1 + {\gamma _k}({x_1})} \right)} $, which can be easily completed via the one-dimensional search over the interval $[0:L - {d_{\min }}]$. We now directly provide the output of Algorithm 1 in this case as $\left\{ {{\mu _k}} \right\}_{k = 1}^K = [\begin{array}{*{20}{c}}
{0.229}&{0.1507}&{0.6203}
\end{array}]$ and $x_{1,\left\{ {{\mu _k}} \right\}_{k = 1}^K}^* = 0 \ \rm{or} \ 6.63\lambda \ \rm{or} \ 15.56\lambda $, under the system parameters of $N = K = 3$, $\left\{ {{\theta _i}} \right\}_{i = 1}^3 = \left[ {\begin{array}{*{20}{c}}
{\pi /7}&{\pi /6.7}&{\pi /6}
\end{array}} \right]$, $\left\{ {{\phi _i}} \right\}_{i = 1}^3 = \left[ {\begin{array}{*{20}{c}}
{\pi /7.2}&{\pi /6.5}&{\pi /5.8}
\end{array}} \right]$, $L = 20\lambda $, ${d_{\min }} = 0.5\lambda $, ${P_k} = 10$ dBm, and ${\beta _k}/{\sigma ^2} = 1$, $k \in {\cal K}$. Subsequently, we present the achievable rate of these three users under different ${x_1} \in [0:19.5\lambda ]$ as shown Fig. 2, from which we can intuitively judge the key reasons why the first track at the BS should employ time-sharing among three different horizontal coordinates ($0$, $6.63\lambda $ and $15.56\lambda $) over the period $T$: A static deployment inevitably favors some users while compromising others, leading to inherent rate unfairness. By periodically switching between these three coordinates for the first track, the system can maximally balance the achievable rate of each user.

After determining these three deployment patterns described above, the optimal time allocation for each of them can be obtained by solving (P3). Taking $T = 100$ s as an example, the outcomes of (P3) are directly given as $t_1^* = 17.551$ s, $t_2^* = 30.7109$ s and $t_3^* = 51.7381$ s, based on which the minimum average rate can be finally computed as 2.4475 bits/s/Hz. In contrast, if the conventional static deployment strategy\footnote{The static deployment strategy means that the BS always adopts a single fixed deployment strategy throughout the entire period. Such a pattern can be determined by replacing the objective in (29) with $\mathop {\min }\limits_{k \in {\cal K}} \left\{ {{{\log }_2}(1 + {\gamma _k}({\bf{x}}))} \right\}$ and then solving the resulting problem.} is exploited for the minimum rate maximization, from Fig. 2 it is not difficult to determine that the optimal static $x_1$ is $6.63\lambda $, yielding the minimum average rate of only 1.9 bits/s/Hz. Therefore, in this case, our proposed strategy achieves the rate gain of 0.5475 bits/s/Hz, representing a performance improvement of approximately $22.37\%$.

  \begin{figure*}[b!]
  \hrulefill
\setcounter{equation}{39}
\begin{equation} \small
\begin{split}{}
{\left[ {{\bf{x}}_{\left\{ {{\mu _k}} \right\}_{k = 1}^K,\pi (i) \to \pi (i + 1)}^t} \right]_m} = \left\{ {\begin{array}{*{20}{c}}
{{{\left[ {{{\bf{x}}_{\left\{ {{\mu _k}} \right\}_{k = 1}^K,\pi (i)}}} \right]}_m} + {1_{\pi (i) \to \pi (i + 1)}}{V_{\max }}t}&{0 \le t \le \frac{{\left| {{{\left[ {{{\bf{x}}_{\left\{ {{\mu _k}} \right\}_{k = 1}^K,\pi (i + 1)}}} \right]}_m} - {{\left[ {{{\bf{x}}_{\left\{ {{\mu _k}} \right\}_{k = 1}^K,\pi (i)}}} \right]}_m}} \right|}}{{{V_{\max }}}}}\\
{{{\left[ {{{\bf{x}}_{\left\{ {{\mu _k}} \right\}_{k = 1}^K,\pi (i + 1)}}} \right]}_m}}&{\frac{{\left| {{{\left[ {{{\bf{x}}_{\left\{ {{\mu _k}} \right\}_{k = 1}^K,\pi (i + 1)}}} \right]}_m} - {{\left[ {{{\bf{x}}_{\left\{ {{\mu _k}} \right\}_{k = 1}^K,\pi (i)}}} \right]}_m}} \right|}}{{{V_{\max }}}} \le t \le {t_{\pi (i)\pi (i + 1)}}}
\end{array}} \right..
\end{split}
\end{equation}
\end{figure*}

\section{Minimum Rate Maximization With Antenna Movement Speed Constraint}
In this section, we investigate the original minimum rate maximization problem (P1) by considering the practical constraint of limited antenna movement speed in (7b). The corresponding problem becomes highly non-convex due to the continuous time variables. To effectively solve it, we propose one suboptimal scheme in the follows, referred to as the Successive Stay-then-Move Trajectory (SSMT).
\subsection{Successive Stay-then-Move Trajectory (SSMT)}
The design of this scheme is primarily motivated by the trajectory optimization details obtained in the ideal case without antenna movement speed constraints. Specifically, in section III, we have derived $\Gamma$ antenna deployment patterns ${{\bf{x}}_{\left\{ {{\mu _k}} \right\}_{k = 1}^K,i}}$ ($i = 1,..,\Gamma $), which should be time-shared by the BS to achieve the optimal performance. Inspired by this, using the SSMT scheme, the BS will sequentially stay at each of these $\Gamma$ deployment patterns for a certain time duration and then switch to the next pattern by utilizing the maximum antenna movement speed. Specifically, to efficiently implement the SSMT scheme, three key problems must be addressed, as outlined below.

1) \textbf{The first problem in the SSMT scheme lies in that how to sort these $\Gamma$ patterns to allow the BS to deploy sequentially, thereby minimizing the overall switching time and then freeing up the longest remaining time for communications at these $\Gamma$ patterns?} To solve this problem, we now derive the time consumed to switch from the pattern ${{{\bf{x}}_{\left\{ {{\mu _k}} \right\}_{k = 1}^K,i}}}$ to the pattern ${{{\bf{x}}_{\left\{ {{\mu _k}} \right\}_{k = 1}^K,j}}}$ as
  \begin{equation}
  \setcounter{equation}{37}
\begin{split}{}
{t_{ij}} = \frac{{\mathop {\max }\limits_{m \in {\cal M}} \left| {{{\left[ {{{\bf{x}}_{\left\{ {{\mu _k}} \right\}_{k = 1}^K,i}}} \right]}_m} - {{\left[ {{{\bf{x}}_{\left\{ {{\mu _k}} \right\}_{k = 1}^K,j}}} \right]}_m}} \right|}}{{{V_{\max }}}},
\end{split}
  \end{equation}
which is essentially the Chebyshev distance between the two deployment patterns normalized by the maximum movement speed ${{V_{\max }}}$. Given the pairwise switching time $\left\{ {{t_{ij}}} \right\}$, the problem of finding the optimal deployment sequence that minimizes the total switching time can be formulated as a shortest Hamiltonian path problem on a complete graph of $\Gamma$ vertices. When $\Gamma$ is small (e.g., $\Gamma \leq 20$), the globally optimal sequence can be obtained exactly via dynamic programming (e.g., the Held-Karp algorithm \cite{TSP_P1}) with the complexity of ${\cal O}(\Gamma^2 2^\Gamma)$, the details of which are omitted here due to its maturity. For the larger $\Gamma$, efficient approximation algorithms such as the Christofides algorithm \cite{TSP_P2} (with a $1.5$-approximation ratio) can be employed to obtain high-quality suboptimal solutions in polynomial time.

Denote the optimal or suboptimal deployment sequence obtained via the Held-Karp algorithm or the Christofides algorithm as ${{\bf{x}}_{\left\{ {{\mu _k}} \right\}_{k = 1}^K,\pi (1)}},...,{{\bf{x}}_{\left\{ {{\mu _k}} \right\}_{k = 1}^K,\pi (\Gamma )}}$, where $\pi (i) \in \left\{ {1,...,\Gamma } \right\},i = 1,...,\Gamma $, and $\pi (i) \ne \pi (j),\forall i \ne j$. In addition, the corresponding total switching time can be derived as
  \begin{equation}
\begin{split}{}
{t_{{\rm{swi}},\left\{ {\pi (i)} \right\}_{i = 1}^\Gamma }} = \sum\nolimits_{i = 1}^{\Gamma  - 1} {{t_{\pi (i)\pi (i + 1)}}}.
\end{split}
  \end{equation}

2) \textbf{After determining the deployment sequence, the second problem in the SSMT scheme lies in that whether there exist signal receiving coupling effects during the switching process from one deployment pattern to another?} More specifically, this problem concerns whether the horizontal distance between any two vertical sliding tracks always remains greater than or equal to ${d_{\min }}$ during the switching process. If the answer is yes, the BS can implement effective signal reception not only when using the $\Gamma $ deployment patterns, but also during all intermediate transitions between them.

\textbf{Lemma 2:} No signal coupling effects occur during the switching process from ${{\bf{x}}_{\left\{ {{\mu _k}} \right\}_{k = 1}^K,\pi (i)}}$ to ${{\bf{x}}_{\left\{ {{\mu _k}} \right\}_{k = 1}^K,\pi (i + 1)}}$, $\forall i = 1,...,\Gamma  - 1$.
\begin{proof}
Please see the Appendix.
\end{proof}

Based on Lemma 2, the obtained rate of user $k$ (in bits/Hz) during the total switching process can be derived as
  \begin{equation}
\begin{split}{}
&{r_{k,{\rm{swi}}}} \\
=& \sum\nolimits_{i = 1}^{\Gamma  - 1} {\int_0^{{t_{\pi (i)\pi (i + 1)}}} {{{\log }_2}(1 + {\gamma _k}({\bf{x}}_{\left\{ {{\mu _k}} \right\}_{k = 1}^K,\pi (i) \to \pi (i + 1)}^t))dt} },
\end{split}
  \end{equation}
where ${{\bf{x}}_{\left\{ {{\mu _k}} \right\}_{k = 1}^K,\pi (i) \to \pi (i + 1)}^t}$ denotes the real-time antenna deployment pattern at time $t$ during the switching process from ${{\bf{x}}_{\left\{ {{\mu _k}} \right\}_{k = 1}^K,\pi (i)}}$ to ${{\bf{x}}_{\left\{ {{\mu _k}} \right\}_{k = 1}^K,\pi (i + 1)}}$, and the $m$-th element in ${\bf{x}}_{\left\{ {{\mu _k}} \right\}_{k = 1}^K,\pi (i) \to \pi (i + 1)}^t$ can be derived as in (40), with
  \begin{equation}
    \setcounter{equation}{41}
  \begin{split}{}
&{1_{\pi (i) \to \pi (i + 1)}}\\
 =& \left\{ {\begin{array}{*{20}{c}}
1&{{{\left[ {{{\bf{x}}_{\left\{ {{\mu _k}} \right\}_{k = 1}^K,\pi (i + 1)}}} \right]}_m} \ge {{\left[ {{{\bf{x}}_{\left\{ {{\mu _k}} \right\}_{k = 1}^K,\pi (i)}}} \right]}_m}}\\
{ - 1}&{{{\left[ {{{\bf{x}}_{\left\{ {{\mu _k}} \right\}_{k = 1}^K,\pi (i + 1)}}} \right]}_m} < {{\left[ {{{\bf{x}}_{\left\{ {{\mu _k}} \right\}_{k = 1}^K,\pi (i)}}} \right]}_m}}
\end{array}} \right..
\end{split}
  \end{equation}

3) \textbf{With ${t_{{\rm{swi}},\left\{ {\pi (i)} \right\}_{i = 1}^\Gamma }}$ and $\left\{ {{r _{k,{\rm{swi}}}}} \right\}_{k = 1}^K$ at hand, the final work is to derive the optimal time duration that the BS stays at each of the above $\Gamma $ deployment patterns, such that the minimum average rate can be maximized.} The corresponding problem can be easily formulated as
 \begin{align}
&({\rm{P4}}):{\rm{  }}\mathop {\max }\limits_{\left\{ {{t_i}} \right\}_{i = 1}^\Gamma , r } \ r  \tag{${\rm{42a}}$}\\
{\rm{              }}&{\rm{s.t.}} \ \ \ \ \frac{1}{T}\left( {\sum\nolimits_{i = 1}^\Gamma  {{t_i}{{\log }_2}(1 + {\gamma _k}({{\bf{x}}_{\left\{ {{\mu _k}} \right\}_{k = 1}^K,i}}))}  + {r _{k,{\rm{swi}}}}} \right) \nonumber\\
&\quad \quad \quad \quad \quad \quad \quad \quad \quad \quad \quad \quad \quad \quad \quad \ \ \ge r  ,\forall k \in {\cal K},\tag{${\rm{42b}}$}\\
 &\ \ \ \ \ \ \ \sum\nolimits_{i = 1}^\Gamma  {{t_i}} = T - {t_{{\rm{swi}},\left\{ {\pi (i)} \right\}_{i = 1}^\Gamma }},\tag{${\rm{42c}}$}
\end{align}
where $\frac{1}{T}\left( {\sum\nolimits_{i = 1}^\Gamma  {{t_i}{{\log }_2}(1 + {\gamma _k}({{\bf{x}}_{\left\{ {{\mu _k}} \right\}_{k = 1}^K,i}}))}  + {r _{k,{\rm{swi}}}}} \right)$ in (42b) is the average rate of user $k$ over the entire period, given the above switching strategy and time allocations $\left\{ {{t_i}} \right\}_{i = 1}^\Gamma $ staying at these $\Gamma $ deployment patterns.

Please note that the above three steps should be implemented only when two conditions are satisfied: i) ${t_{{\rm{swi}},\left\{ {\pi (i)} \right\}_{i = 1}^\Gamma }} < T$; ii) The objective of (P4) is no smaller than the minimum average rate obtained by adopting the static deployment strategy. Actually, the condition i) ensures that the BS can feasibly stay at each of these $\Gamma $ deployment patterns for a certain duration, while the condition ii) guarantees that the proposed dynamic switching strategy outperforms the static deployment, thereby justifying its practical advantage. If either condition fails to hold, the BS will simply adopt the static deployment strategy.

\textbf{Proposition 1:} When $T \to 0$ or ${V_{\max }} \to 0$, the static deployment strategy is optimal in the SSMT scheme. Conversely, if $T \to \infty $ or ${V_{\max }} \to \infty $, the dynamic switching strategy becomes optimal, and its performance approaches the upper bound obtained by solving (P3).

\begin{proof}
When $T \to 0$ or ${V_{\max }} \to 0$, there is almost no extra time for antennas to move (when $T \to 0$), or antenna movements will consume enough time (when ${V_{\max }} \to 0$). In both cases, it is most reasonable for the BS to always stay at the optimal static deployment pattern. When $T \to \infty $ or ${V_{\max }} \to \infty $, the time required for the BS to switch between any two deployment patterns ${{{\bf{x}}_{\left\{ {{\mu _k}} \right\}_{k = 1}^K,i}}}$ and ${{{\bf{x}}_{\left\{ {{\mu _k}} \right\}_{k = 1}^K,j}}}$ (for any $i \ne j$) becomes negligible. Consequently, the BS can allocate time among the $\Gamma$ deployment patterns ${{{\bf{x}}_{\left\{ {{\mu _k}} \right\}_{k = 1}^K,i}}}$ ($i = 1,\ldots,\Gamma$) in a nearly optimal manner.
\end{proof}

\textbf{Complexity Analysis:} In the first problem above, determining the optimal deployment sequence requires the complexity of ${\cal O}(\Gamma^2 2^\Gamma)$. Additionally, the complexity of solving (P4) is about ${\cal O}({(\Gamma  + 1)^{3.5}})$. Therefore, the complexity of the SSMT scheme is about ${\cal O}({\Gamma ^2}{2^\Gamma } + {(\Gamma  + 1)^{3.5}})$, under the premise that these $\Gamma $ deployment patterns are already known.

 \begin{figure} [!t]
\centering
\includegraphics[width=8cm]{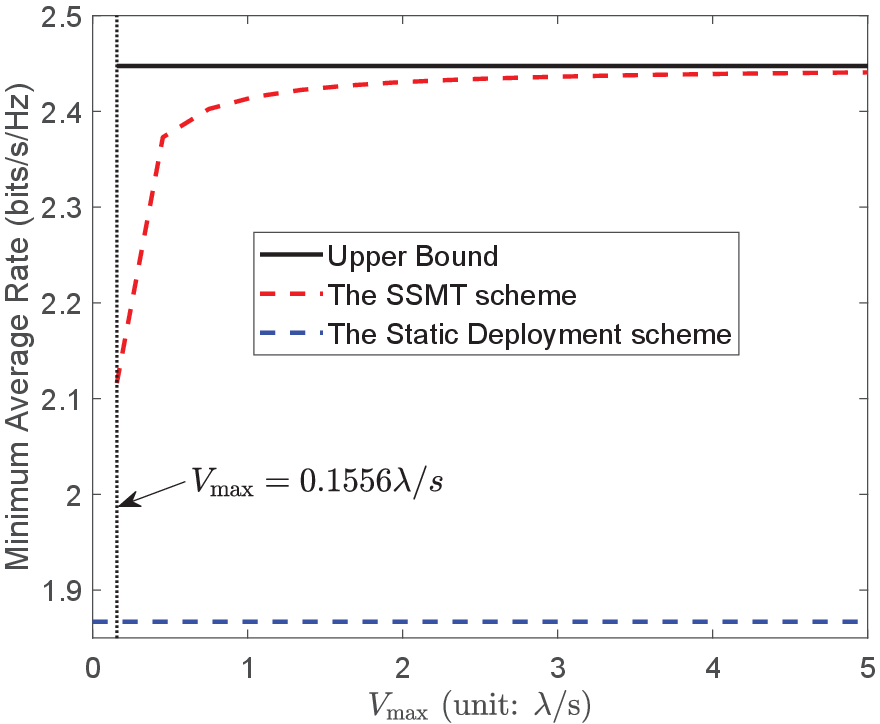}
\captionsetup{font=small}
\caption{The minimum average rate of the SSMT scheme w.r.t. ${V_{\max }}$.} \label{fig:Fig1}
\end{figure}

\subsection{The Special Case of $M = 2$}
We consider the same parameter settings as mentioned in subsection B of section III. Then, with the SSMT scheme, it is not difficult to determine the optimal deployment sequence as ${x_{1,\left\{ {{\mu _k}} \right\}_{k = 1}^K,\pi (1)}} = 0,{x_{1,\left\{ {{\mu _k}} \right\}_{k = 1}^K,\pi (2)}} = 6.63\lambda ,{x_{1,\left\{ {{\mu _k}} \right\}_{k = 1}^K,\pi (3)}} = 15.56\lambda $.\footnote{Or equivalently, ${x_{1,\left\{ {{\mu _k}} \right\}_{k = 1}^K,\pi (1)}} = 15.56\lambda ,{x_{1,\left\{ {{\mu _k}} \right\}_{k = 1}^K,\pi (2)}} = 6.63\lambda ,{x_{1,\left\{ {{\mu _k}} \right\}_{k = 1}^K,\pi (3)}} = 0$.} Therefore, ${t_{{\rm{swi}},\left\{ {\pi (i)} \right\}_{i = 1}^\Gamma }}$ in this case can be derived as
  \begin{equation} \nonumber
\begin{split}{}
{t_{{\rm{swi}},\left\{ {\pi (i)} \right\}_{i = 1}^\Gamma }} &= {t_{\pi (1)\pi (2)}} + {t_{\pi (2)\pi (3)}}\\
 &= \frac{{6.63\lambda }}{{{V_{\max }}}} + \frac{{15.56\lambda  - 6.63\lambda }}{{{V_{\max }}}} = \frac{{15.56\lambda }}{{{V_{\max }}}}.
\end{split}
  \end{equation}

In addition, based on (39)-(41), ${r _{k,{\rm{swi}}}}$ in this case can be easily derived as
  \begin{equation} \nonumber
  \begin{split}{}
{r_{k,{\rm{swi}}}} =& \int_0^{{t_{\pi (1)\pi (2)}}} {{{\log }_2}\left( {1 + {\gamma _k}(x_{1,\left\{ {{\mu _k}} \right\}_{k = 1}^K,\pi (1) \to \pi (2)}^t} \right)dt} \\
 &+ \int_0^{{t_{\pi (2)\pi (3)}}} {{{\log }_2}\left( {1 + {\gamma _k}(x_{1,\left\{ {{\mu _k}} \right\}_{k = 1}^K,\pi (2) \to \pi (3)}^t} \right)dt} \\
 =& \int_0^{\frac{{6.63\lambda }}{{{V_{\max }}}}} {{{\log }_2}(1 + {\gamma _k}({V_{\max }}t))dt} \\
 &+ \int_0^{\frac{{15.56\lambda  - 6.63\lambda }}{{{V_{\max }}}}} {{{\log }_2}(1 + {\gamma _k}(6.63\lambda  + {V_{\max }}t))dt} .
\end{split}
  \end{equation}


Based on the known ${t_{{\rm{swi}},\left\{ {\pi (i)} \right\}_{i = 1}^\Gamma }}$ and $\left\{ {{r _{k,{\rm{swi}}}}} \right\}_{k = 1}^K$, along with the scheme design described above (which includes determining whether $T > {t_{{\rm{swi}},\left\{ {\pi (i)} \right\}_{i = 1}^3}}$; if not, the BS adopts the static deployment strategy; otherwise, the problem (P4) should be solved for obtaining the gain of the dynamic switching strategy and then comparing it with that of the static strategy), the minimum average rate of the SSMT scheme can be derived.

Fig. 3 illustrates the minimum average rate performance of the SSMT scheme w.r.t. ${V_{\max }}$, under the parameter settings provided in subsection B of section III, from which we can obviously observe that: i) When ${V_{\max }}$ lies in the small regime, i.e., $[0,0.1556\lambda ]$, the total switching time exceeds the total duration, leaving no time for the BS to deploy at each of these $\Gamma $ patterns. In this regime, it is preferable for the BS to adopt the static deployment scheme; ii) When ${V_{\max }}$ continues to increase, the switching time decreases proportionally. This allows the BS to spend a larger fraction of the period $T$ actually residing at the $\Gamma $ deployment patterns. The minimum average rate therefore rises monotonically w.r.t. ${V_{\max }}$, reflecting the growing benefit of dynamic time-sharing. For sufficiently high speed, the performance curve saturates and approaches the upper bound provided by the ideal case, thereby validating Proposition 1, i.e. the dynamic switching strategy becomes optimal when ${V_{\max }} \to \infty $.

\begin{table*}[t]
\caption{The Optimal Deployment Patterns of The BS w.r.t. $L$ Without Antenna Movement Speed Constraint.}
\renewcommand\arraystretch{1.35}
\normalsize
\centering
\begin{tabular}[l]{|c|c|c|c|c|c|c|c|c|c|c|c|c|c|}
 \hline
&Optimal Deployment Patterns that Need Time-Sharing\\
 \hline
 $L = 14\lambda $ &Pattern 1: ${{\bf{x}}^*} = [0,L]$; Pattern 2: ${{\bf{x}}^*} = [9.8685\lambda ,L]$\\
 \hline
 $L = 16\lambda $ &Pattern 1: ${{\bf{x}}^*} = [0,L]$; Pattern 2: ${{\bf{x}}^*} = [2.418\lambda ,L]$; Pattern 3: ${{\bf{x}}^*} = [11.532\lambda ,L]$\\
 \hline
  $L = 18\lambda $ &Pattern 1: ${{\bf{x}}^*} = [0,L]$; Pattern 2: ${{\bf{x}}^*} = [4.6025\lambda ,L]$; Pattern 3: ${{\bf{x}}^*} = [13.4575\lambda ,L]$\\
   \hline
  $L = 20\lambda $ &Pattern 1: ${{\bf{x}}^*} = [0,L]$; Pattern 2: ${{\bf{x}}^*} = [6.6325\lambda ,L]$; Pattern 3: ${{\bf{x}}^*} = [15.5615\lambda ,L]$\\
     \hline
  $L = 22\lambda $ &Pattern 1: ${{\bf{x}}^*} = [1.1395\lambda,L]$; Pattern 2: ${{\bf{x}}^*} = [8.686\lambda ,L]$; Pattern 3: ${{\bf{x}}^*} = [17.3935\lambda ,L]$\\
  \hline
 \end{tabular}
{
 \label{tab:tb1}
}
\end{table*}

\section{Simulation Results}
This section provides numerical results to verify the effectiveness of our proposed schemes. Without otherwise stated, the number of users is set as $K = 3$, these users are distributed on a circle with the radius of $r = 10^3$ m centered on the BS, and the corresponding elevation and azimuth AoAs are provided in subsection B of section III. Therefore, the large-scale fading coefficient can be derived as ${\beta _k} = {\beta _0}{r^{ - {\alpha _0}}}$, $\forall k$, where ${\beta _0} = {10^{ - 3}}$ is the reference average channel power gain at 1 m and ${\alpha _0} = 2$ is the path loss exponent. The transmit power of each user is $P_k = 10$ dBm, $\forall k$. For the BS, the minimum distance between any two vertical sliding tracks for avoiding coupling effects is ${d_{\min }} = 0.5\lambda $, and the number of antennas on each vertical sliding track is set as $N = 3$. The noise power is set as ${\sigma ^2} =  - 90$ dBm. The total transmission period is $T = 100$ s. Other simulation-specific parameters will be reflected in the corresponding figure. Furthermore, it should be noted that the following figures compare only three schemes: i) The ideal case with no antenna movement speed constraint (which achieves the performance upper bound); ii) The SSMT scheme; and iii) The static deployment baseline described in Footnote 3. Other benchmarks such as random antenna positions \cite{Lipeng2}, FPA arrays \cite{MA_wideband1}, or antenna selection \cite{MA_MIMO1}, are omitted here as their performance is consistently outperformed by the static deployment baseline.

 \begin{figure} [!t]
\centering
\includegraphics[width=8cm]{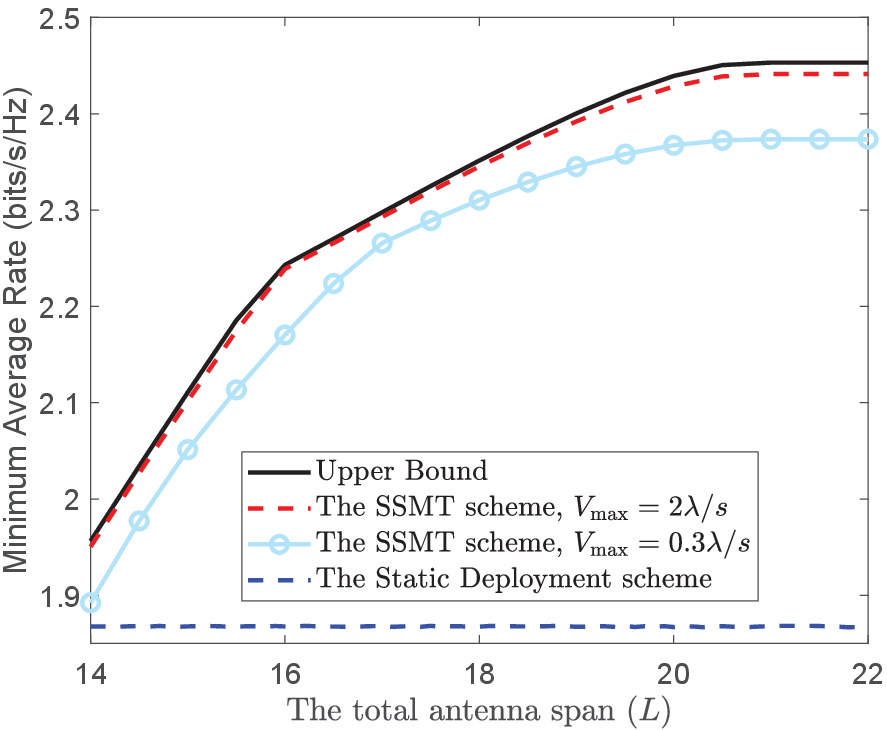}
\captionsetup{font=small}
\caption{The minimum average rate of three different schemes w.r.t. the total antenna span $L$.} \label{fig:Fig1}
\end{figure}

Fig. 4 presents the minimum average rate of three schemes w.r.t. the total antenna span $L$ at the BS given $M = 2$. We can observe that: i) As $L$ increases, the performance of the case without antenna movement speed constraint and the SSMT scheme improves initially but ultimately saturates. This behavior can be explained by the fact that a larger $L$ affords the vertical sliding tracks greater flexibility to adjust their trajectory, thereby enhancing the spatial DoFs and improving the channel conditions. However, as $L$ is further increased, e.g., from $21\lambda $ to $22\lambda $, the spatial DoFs are nearly exhausted, suggesting that it is not necessary to infinitely expand $L$; ii) For the SSMT scheme, as the maximum movement speed of the vertical sliding tracks increases, the time spent on the process of switching among the $\Gamma$ deployment patterns decreases. This allows more time to be allocated to each of the $\Gamma$ patterns. Consequently, the performance of the SSMT scheme gradually approaches the upper bound w.r.t. ${V_{\max}}$. For instance, when ${V_{\max }} = 2\lambda /s$, there is almost no performance gap between the upper bound and the SSMT scheme; iii) For the static deployment scheme, given arbitrary $L$, since only one deployment pattern (even this pattern is carefully optimized) is adopted across the entire period, it cannot mitigate rate unfairness through time sharing, hence resulting in poor performance.

\begin{figure*}[t] 
  \centering
  \includegraphics[width=5.8cm]{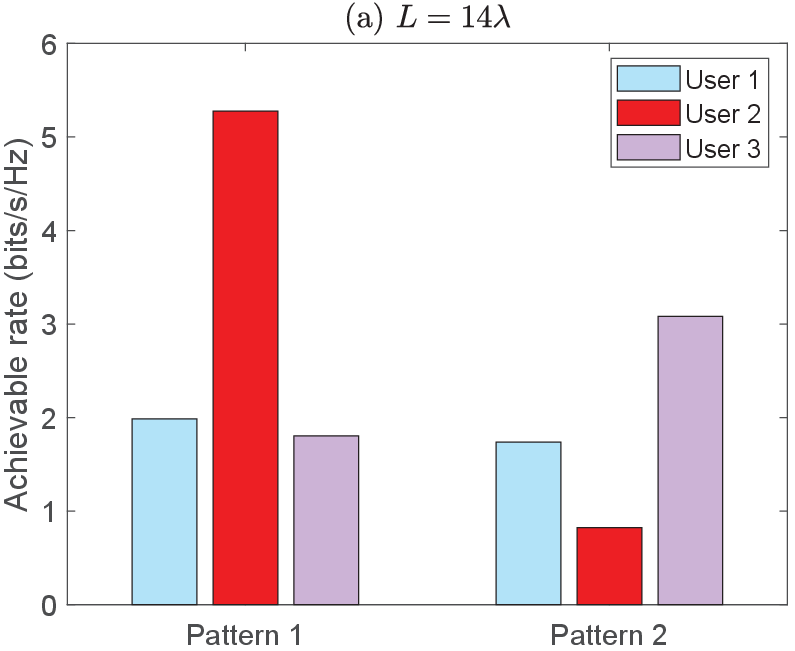}
  \includegraphics[width=5.8cm]{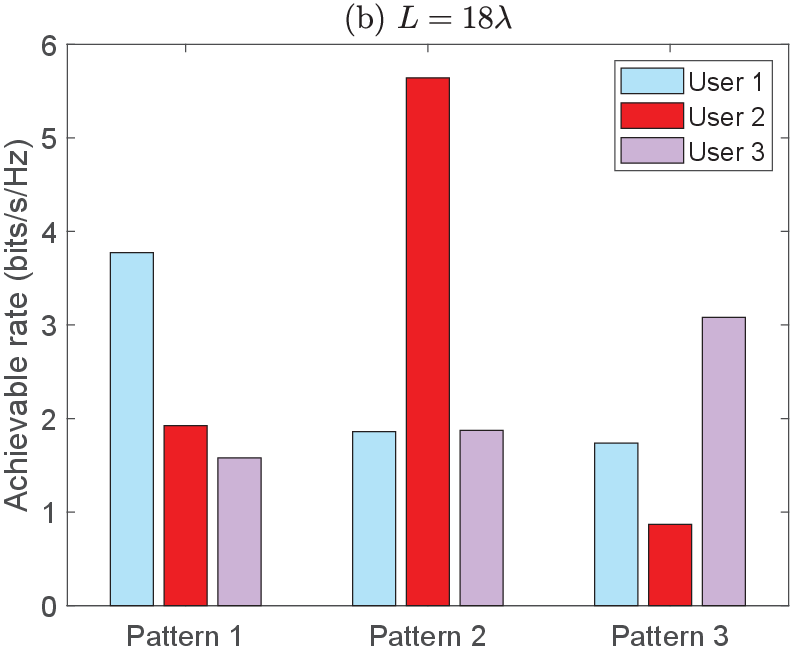}
  \includegraphics[width=5.8cm]{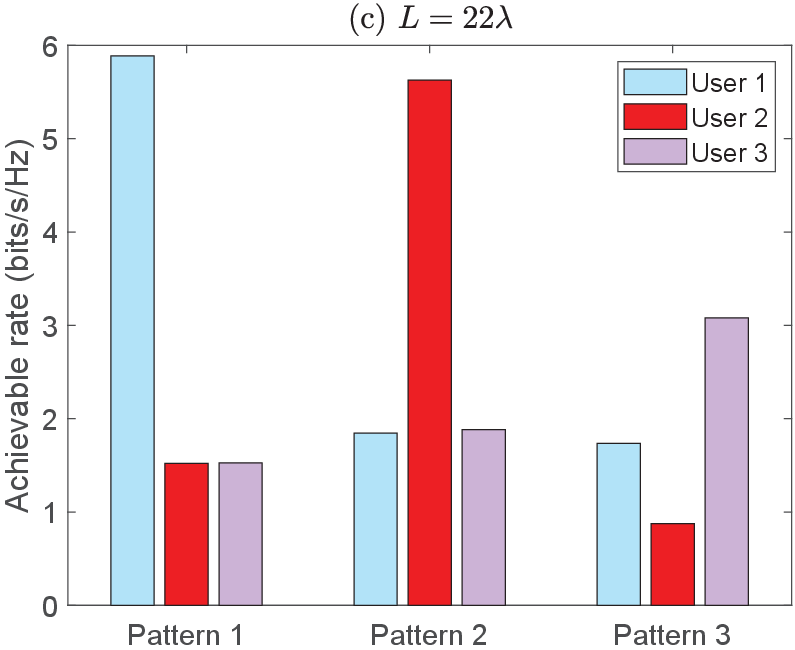}
  \captionsetup{font=small}
  \caption{The case without antenna movement speed constraint: Achievable rate of different users under each of the optimal deployment patterns.}
\end{figure*}

Corresponding to Fig. 4, Table I provides the optimal deployment patterns w.r.t. $L$ in the case without antenna movement speed constraint, Fig. 5 illustrates the achievable rate of different users under each of these optimal deployment patterns, while Fig. 6 further presents the optimal time allocations under each of the optimal deployment patterns. Specifically,
\begin{itemize}
\item[$\bullet$]  First, as shown in Table I, the number of optimal deployment patterns gradually increases w.r.t. $L$. For instance, there are two patterns when $L = 14\lambda $ and three patterns when $L = 16\lambda $, $L = 18\lambda $, $L = 20\lambda $ and $L = 22\lambda $.
    This pattern growth is intuitive, since a larger antenna span provides greater flexibility for spatial reconfiguration, allowing the system to exploit a richer set of deployment layouts. More patterns imply more opportunities to create favorable channel conditions for different users at different times, which is essential for balancing the achievable rate. Furthermore, as $L$ increases, the distance between the two vertical sliding tracks widens in each pattern. This sparser configuration provides better conditions for the BS to effectively differentiate and isolate signals from different users.
\item[$\bullet$] Second, as presented in Fig. 5, for arbitrary $L = 14\lambda $, $L = 18\lambda $ or $L = 20\lambda $, a key observation is that each user achieves its highest rate under only one specific deployment pattern, and the pattern that yields the best performance differs across users. This clearly indicates that no single static antenna layout can simultaneously optimize the channel conditions for all users, i.e., any fixed deployment inevitably favors some users while disadvantaging others, leading to inherent rate unfairness. This user-pattern heterogeneity fundamentally motivates the necessity of time-sharing strategies, i.e., by dynamically switching among multiple deployment patterns over time, the BS can balance the achievable rate across all users, thereby improving system fairness.

\item[$\bullet$] Third, as shown in Fig. 6, the optimal time allocations for the deployment patterns vary significantly w.r.t. $L$. When $L$ is small (e.g., $14\lambda $), the transmission time is concentrated on a single dominant pattern. However, as $L$ increases (e.g., $18\lambda $ and $20\lambda $), the time is distributed more uniformly across multiple patterns. This evolution demonstrates that a larger antenna span enhances the system's ability to exploit time-sharing flexibility, thereby enabling a more equitable distribution of channel resources and better achieving the objective of rate fairness.
\end{itemize}

 \begin{figure} [!t]
\centering
\includegraphics[width=8cm]{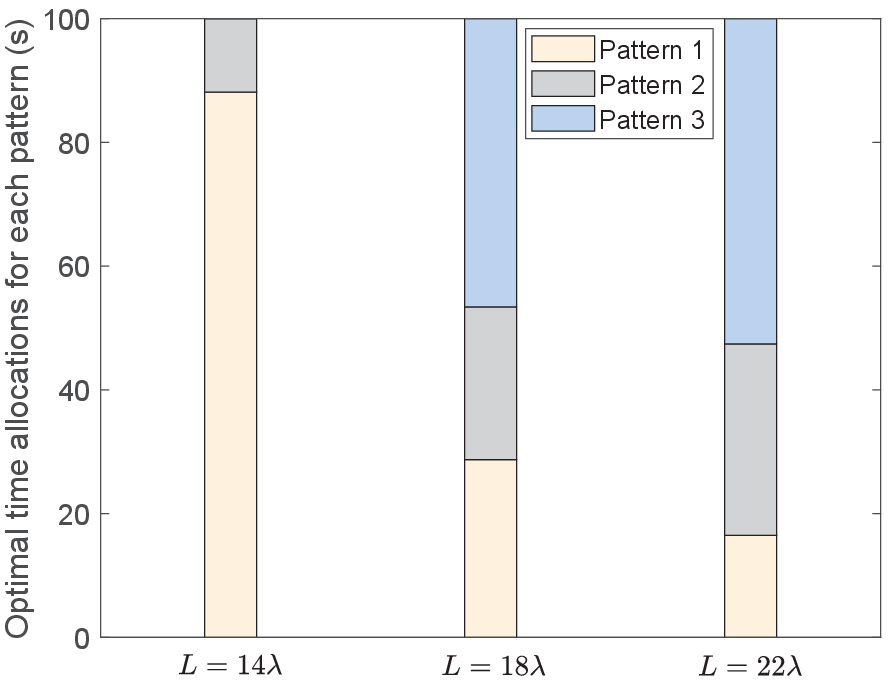}
\captionsetup{font=small}
\caption{The case without antenna movement speed constraint: The optimal time allocations for each pattern w.r.t. different $L$.} \label{fig:Fig1}
\end{figure}

\begin{figure} [!t]
\centering
\includegraphics[width=8cm]{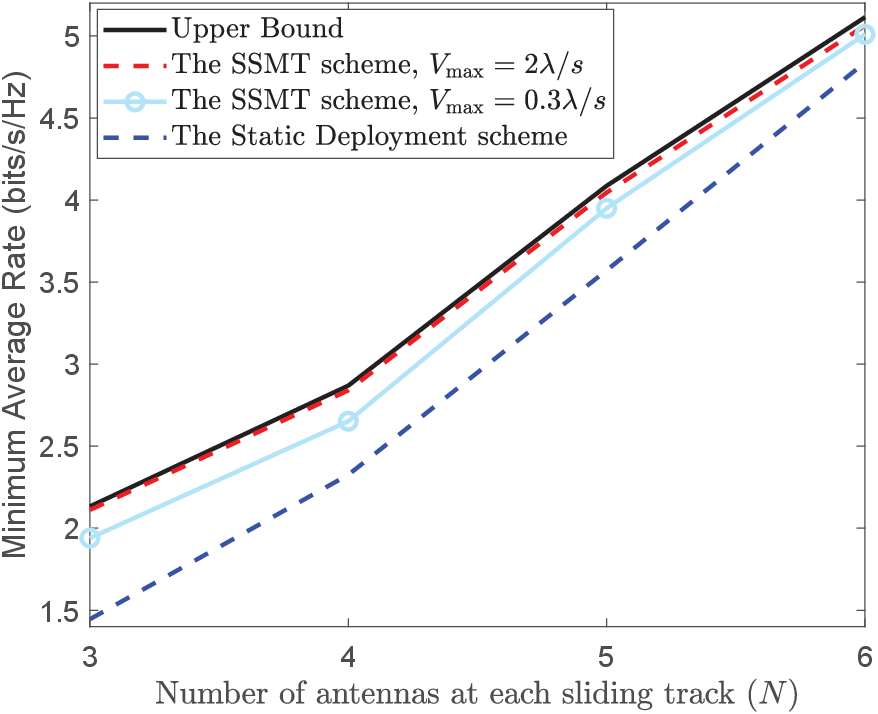}
\captionsetup{font=small}
\caption{The minimum average rate of the three different schemes w.r.t. the number of antennas $N$ at each sliding track.} \label{fig:Fig1}
\end{figure}

Fig. 7 illustrates the minimum average rate w.r.t. the number of antennas $N$ at each sliding track given $M = 2$ and $L = 20\lambda $, from which we can observe that: i) For all three schemes, the minimum average rate increases monotonically w.r.t. $N$. This is expected because a larger $N$ provides higher beamforming resolution and diversity gain, thereby improving the SINR for each user; ii) The performance gap between the ideal case (without movement speed constraint) and the static deployment scheme is most pronounced when $N$ is small. As $N$ increases, this gap gradually shrinks. This trend suggests that when the array is already large, the additional spatial flexibility offered by time-sharing contributes less to the overall performance, as the system can rely on the inherent spatial diversity of a large static array; iii) The SSMT scheme consistently bridges the gap between the ideal upper bound and the static baseline. Even with a moderate $N$, the SSMT scheme achieves a significant portion of the ideal performance, demonstrating its practical value when antenna movement speed is limited.

\begin{figure} [!t]
\centering
\includegraphics[width=8cm]{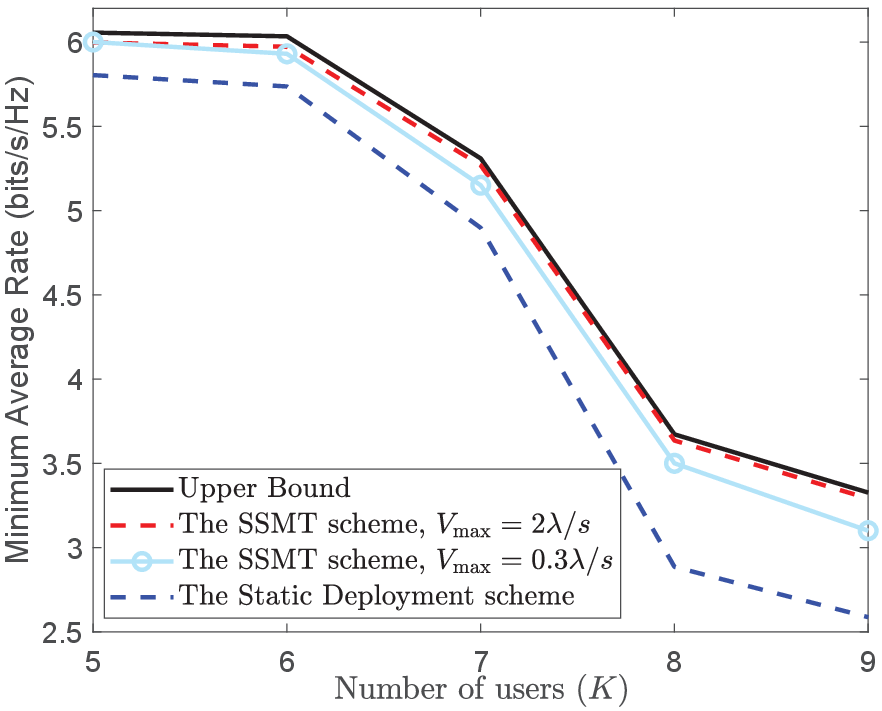}
\captionsetup{font=small}
\caption{The minimum average rate of the three different schemes w.r.t. the number of users $K$.} \label{fig:Fig1}
\end{figure}

Furthermore, Fig. 8 presents the minimum average rate w.r.t. the number of users $K$ given $M = 3$ and $L = 20\lambda $, where for arbitrary $K \in [5,9]$, the physical elevation and azimuth AoAs of user $k$ are set as ${\theta _k} = {\left[ {\bf{\Theta }} \right]_k}$ and ${\phi _k} = {\left[ {\bf{\Phi }} \right]_k}$, with
  \begin{equation} \nonumber
  \begin{split}{}
{\bf{\Theta }} &= [1.41 \ 1.14 \ 1.81 \ 0.18 \ 3.12 \ 2.91 \ 0.73 \ 1.09 \ 2.98]\\
{\bf{\Phi }} &= [0.72 \ 0.69 \ 1.65 \ 2.23 \ 2.28 \ 0.41 \ 1.62 \ 0.59 \ 0.38].
 \end{split}
 \end{equation}
From Fig. 8 we can observe that: i) A general declining trend appears for all schemes w.r.t. $K$. This is because that for the BS, serving more users over the same time-frequency resources intensifies inter-user interference, thereby reducing the achievable rate per user; ii) The static deployment scheme suffers the most significant performance drop. With only a single and fixed antenna pattern, it cannot adapt to the differentiated channel conditions of different users, leading to severe unfairness and a low minimum rate; iii) In contrast, both the ideal case and the SSMT scheme show a slower performance decay. Their ability to dynamically adjust antenna positions and allocate time across multiple deployment patterns allows them to better mitigate the negative impact of increased multi-user interference. This demonstrates the essential advantage of the time-sharing strategy in scalable multiuser systems.


\begin{figure} [!t]
\centering
\includegraphics[width=8cm]{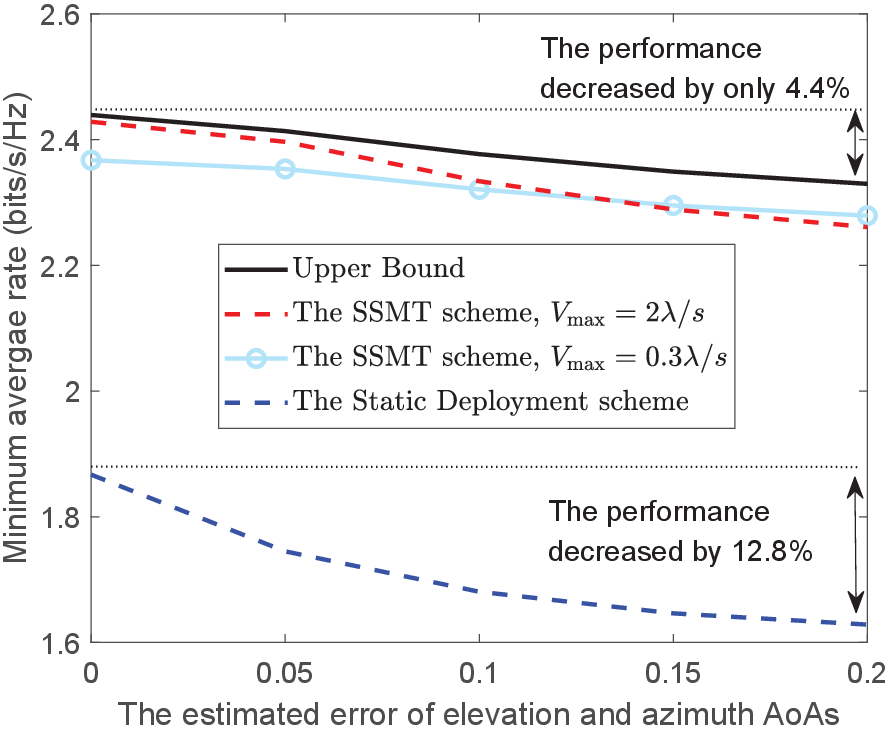}
\captionsetup{font=small}
\caption{The minimum average rate of the three different schemes w.r.t. the angle estimation error.} \label{fig:Fig1}
\end{figure}

Finally, the robustness of the proposed schemes is assessed in Fig. 9 given $M = 2$ and $L = 20\lambda $, which shows the minimum average rate under different levels of angle estimation error. Here, such error indicates that the BS can only obtain the inaccurate versions (denoted as $\widetilde {{\theta _k}}$ and $\widetilde {{\phi _k}}$) of the actual elevation and azimuth AoAs for user $k$, $\forall k$. More specifically, $\widetilde {{\theta _k}} = {\theta _k} + {\rm{error}}$ and $\widetilde {{\phi _k}} = {\phi _k} + {\rm{error}}$. The BS will rely on the inaccurate AoAs to perform optimization designs, while the results of Fig. 9 are computed by applying optimization results to actual AoAs. From Fig. 9 we can observe that: i) Imperfect knowledge of the users' AoAs negatively impacts all schemes. The performance degradation stems from mismatches between the assumed and actual channel directions, leading to suboptimal antenna movement trajectory and concurrently the reduced beamforming gain; ii) The static deployment scheme is the most vulnerable to estimation errors. Its fixed pattern cannot compensate for the channel mismatch, resulting in a rapid performance drop; iii) The schemes (both the ideal case and the SSMT scheme) leveraging the time-sharing strategy show greater resilience. Their ability to reconfigure antenna positions over time provides a form of spatial adaptation that can partially counteract the effects of inaccurate channel information. This indicates a key practical benefit of time-sharing-enabled MA systems: They are not only capable of optimizing performance under perfect CSI but also maintaining reasonable performance in the presence of channel uncertainties.

\section{Conclusions}
This paper investigates the minimum rate maximization problem in MAs-aided multiuser uplink systems by designing adaptive antenna trajectory at the BS. The theoretical analysis under ideal unlimited antenna movement speed conditions reveals that the optimal strategy only requires time-sharing among a finite set of antenna deployment patterns, avoiding frequent movements and providing an energy-efficient guideline for practical MA systems. For the practical scenario with limited movement speed, we propose a heuristic SSMT scheme, which achieves a favorable performance-complexity trade-off while ensuring no signal coupling during transitions. Simulation results validate the substantial gains of the proposed designs over the conventional static deployment, and further indicate that: i) The rate performance saturates as the antenna movement range increases, and ii) Time-sharing is essential to mitigate user channel heterogeneity and enhance fairness. This work provides a practical trajectory design framework and valuable insights for fairness-oriented resource scheduling in MAs-enabled multiuser communication systems.

\begin{appendix}
\end{appendix}

\section{Proof of Lemma 2}
 During the switching process from the pattern ${{\bf{x}}_{\left\{ {{\mu _k}} \right\}_{k = 1}^K,\pi (i)}}$ to ${{\bf{x}}_{\left\{ {{\mu _k}} \right\}_{k = 1}^K,\pi (i + 1)}}$, we denote the distance between ${\left[ {\bf{x}} \right]_m}$ and ${\left[ {\bf{x}} \right]_{m - 1}}$ at time $t$ ($t \in [0,{t_{\pi (i)\pi (i + 1)}}]$) as
\begin{equation}
 \setcounter{equation}{43}
{{\rm{D}}_m}(t) = {\left[ {{{\bf{x}}^t}} \right]_m} - {\left[ {{{\bf{x}}^t}} \right]_{m - 1}}.
\end{equation}
Since the movement speed of each antenna is the constant ${V_{\max }}$ with two possible directions, hence
\begin{equation}
\begin{split}{}
&d\left[ {{{\rm{D}}_m}(t)} \right]/dt\\
 =& d{\left[ {{{\bf{x}}^t}} \right]_m}/dt - d{\left[ {{{\bf{x}}^t}} \right]_{m - 1}}/dt \in \left\{ { - 2{V_{\max }},0,2{V_{\max }}} \right\}.
\end{split}
\end{equation}

Therefore, ${{\rm{D}}_m}(t)$ is an affine function w.r.t. $t$, i.e.,
\begin{equation}
\begin{split}{}
&{{\rm{D}}_m}(t)\\
 =& {{\rm{D}}_m}(0) + d\left[ {{{\rm{D}}_m}(t)} \right]/dt \times t,t \in [0,{t_{\pi (i)\pi (i + 1)}}],
\end{split}
\end{equation}
where
\begin{equation} \nonumber
\begin{split}{}
{{\rm{D}}_m}(0) =& {\left[ {{{\bf{x}}_{\left\{ {{\mu _k}} \right\}_{k = 1}^K,\pi (i)}}} \right]_m} - {\left[ {{{\bf{x}}_{\left\{ {{\mu _k}} \right\}_{k = 1}^K,\pi (i)}}} \right]_{m - 1}},\\
{{\rm{D}}_m}({t_{\pi (i)\pi (i + 1)}}) =& {\left[ {{{\bf{x}}_{\left\{ {{\mu _k}} \right\}_{k = 1}^K,\pi (i + 1)}}} \right]_m} - {\left[ {{{\bf{x}}_{\left\{ {{\mu _k}} \right\}_{k = 1}^K,\pi (i + 1)}}} \right]_{m - 1}}.
\end{split}
\end{equation}

Note that ${{\rm{D}}_m}(0) \ge {d_{\min }}$ and ${{\rm{D}}_m}({t_{\pi (i)\pi (i + 1)}}) \ge {d_{\min }}$, $\forall m$. Then, we examine three cases according to the value of $d\left[ {{{\rm{D}}_m}(t)} \right]/dt$ as follows.
\begin{itemize}
\item[$\bullet$]  Case 1: $d\left[ {{{\rm{D}}_m}(t)} \right]/dt =  - 2{V_{\max }}$. In this case, ${{\rm{D}}_m}(t)$ is strictly decreasing w.r.t. $t \in [0,{t_{\pi (i)\pi (i + 1)}}]$. Therefore,
     \begin{equation} \nonumber
\begin{split}{}
{{\rm{D}}_m}(t) \ge {{\rm{D}}_m}({t_{\pi (i)\pi (i + 1)}}) \ge {d_{\min }},t \in [0,{t_{\pi (i)\pi (i + 1)}}].
\end{split}
\end{equation}

\item[$\bullet$]  Case 2: $d\left[ {{{\rm{D}}_m}(t)} \right]/dt =  0$. In this case, ${{\rm{D}}_m}(t)$ remains the constant w.r.t. $t \in [0,{t_{\pi (i)\pi (i + 1)}}]$, i.e.,
     \begin{equation} \nonumber
\begin{split}{}
{{\rm{D}}_m}(t) = {{\rm{D}}_m}(0) \ge {d_{\min }},t \in [0,{t_{\pi (i)\pi (i + 1)}}].
\end{split}
\end{equation}

\item[$\bullet$]  Case 3: $d\left[ {{{\rm{D}}_m}(t)} \right]/dt =  2{V_{\max }}$. In this case, ${{\rm{D}}_m}(t)$ is strictly increasing w.r.t. $t \in [0,{t_{\pi (i)\pi (i + 1)}}]$. Therefore,
     \begin{equation} \nonumber
\begin{split}{}
{{\rm{D}}_m}(t) \ge {{\rm{D}}_m}(0) \ge {d_{\min }},t \in [0,{t_{\pi (i)\pi (i + 1)}}].
\end{split}
\end{equation}
\end{itemize}

In every possible case we obtain that
     \begin{equation}
\begin{split}{}
\mathop {\min }\limits_{t \in [0,{t_{\pi (i)\pi (i + 1)}}]} {{\rm{D}}_m}(t) \ge {d_{\min }}, \forall m.
\end{split}
\end{equation}
This completes the proof.

\normalem
\bibliographystyle{IEEEtran}
\bibliography{IEEEabrv,mybib}

\end{document}